\documentclass[aps,twocolumn,showpacs,prb,reprint,amsmath,longbibliography]{revtex4-2}
            
\usepackage{bbm}
\usepackage{mathrsfs}
\usepackage{amsmath}
\usepackage{amsfonts}
\usepackage[colorlinks=true, 
            citecolor=blue, 
            linkcolor=blue, 
            urlcolor=blue]{hyperref}
\usepackage{graphicx,epstopdf}
\usepackage{subfigure}
\usepackage{epsfig}
\usepackage{dcolumn}
\usepackage{bm}
\usepackage{color}
\usepackage{natbib}
\usepackage{amssymb}  
\usepackage{xcolor}
\usepackage{braket}
     
\begin{document}

\title{Three Magnetization Peaks in HgBa$_2$Ca$_2$Cu$_3$O$_8$ Single Crystals}

\author{Yongze Ye$^1$}
\author{Yuhao Liu$^1$}
\author{Wenshan Hong$^2$}
\author{Yuan Li$^2$}
\author{Hai-Hu Wen$^1$}\email{hhwen@nju.edu.cn}

\affiliation{
$^1$Center for Superconducting Physics and Materials, National Laboratory of Solid State Microstructures and Department of Physics, Collaborative Innovation Center for Advanced Microstructures, Nanjing University, Nanjing 210093, China\\
$^2$International Center for Quantum Materials, School of Physics, Peking University, Beijing 100871, China
}

\date{\today}

\begin{abstract}
By measuring magnetization hysteresis loops of the superconducting HgBa$_2$Ca$_2$Cu$_3$O$_8$ single crystals  ($T_{\rm c}$ = 133 K), we observed three magnetization peaks in a wide temperature region. This is in contrast to the previous observation that there are only two magnetization peaks in many superconductors. Detailed analysis finds that the second peak here evolves from a kinky structure at low temperatures and gets enhanced at high temperatures; the third peak evolves from a general broad peak at low temperatures and evolves into a sharp peak and even a step-like one at high temperatures. We propose a general phase diagram to interpret these peaks, the second peak is corresponding to the order-disorder transition, while the third peak is associated with the elastic-plastic crossover. Our work unifies the understanding of different ``second peak'' structures in different systems and thus sheds new light in understanding the vortex dynamics in type-II superconductors.
\end{abstract}
\pacs{74.25.Dw, 74.25.Op, 74.25.Qt, 74.25.Uv} 

\maketitle

Due to the high superconducting transition temperature, strong anisotropy, and short coherence length, the mixed state of cuprate high temperature superconductors (HTSCs) possesses very interesting and rich physics. Extensive investigations have been conducted to explore the properties of vortex state of high-temperature superconductors. Because of the great experimental and theoretical efforts, various vortex states have been distinguished, making vortex phase diagram more complex and intriguing\cite{Blatter1994,VINOKUR1998,klein2001bragg,avraham2001inverse,beidenkopf2005equilibrium}. One of the important discoveries is the second peak (SP) effect\cite{LeBlancandLittle1961,KOPYLOV1990291,kokkaliaris2000transition} on the magnetization hysteresis loops (MHLs), which exhibits an enhancement of critical current density with applied magnetic field after the first peak near zero field. Thus this SP effect may be very useful for applications. The intimate reason for the SP effect is still under debate because it exhibits different forms in different systems\cite{Banerjee2000Peak,Yang2008PE,Bonura2012PE,chowdhury2003peak}. One common consensus is that the SP effect of magnetization should reflect the transition between different vortex states. In most previous experiments, people only observed two peaks in a variety of type-II superconductors, including both conventional and high temperature superconductors.\par

The SP effect peak was first found in a conventional superconductor 2H-NbSe$_2$. This kind of SP is very sharp and locates near the upper critical field $H_{\rm c2}$\cite{Ghosh1996peak,Banerjee1998peak}. It was considered to be related to the softening of the vortex lattice\cite{Bhattacharya1993disorder,Troyanovski2002STM}, which was first proposed by Pippard\cite{pippard1969PE}, and subsequently elaborated by Larkin and Ovchinnikov\cite{larkin1979pinning}. In cuprate HTSCs, The SP effect can be broadly classified into two types: one is found in Bi$_2$Sr$_2$CaCu$_2$O$_8$ (Bi-2212), and the other in YBa$_2$Cu$_3$O$_7$(YBCO). In Bi-2212, SP effect occurs in the field region of 200-700 Oe depending on the disorder and doping status of the samples, and the magnetic field associating with the second peak is almost independent of temperature\cite{Khaykovich1996VL,li2002bg}, usually in the temperature region between 20 to 50 K. The magnetization rises sharply to form a second peak with increasing magnetic field, then falls gently, exhibiting a cliff-like feature. Experiments, such as Small Angle Neutron Scattering (SANS), demonstrated that the second peak effect in Bi-2212 corresponds to the order-disorder transition from Bragg glass to vortex glass.\cite{cubitt1993direct}.\par

In contrast to Bi-2212, the second peak found in YBCO shows a temperature dependent field position, and sometimes the field for the SP exhibits a non-monotonic temperature dependence\cite{Nishizaki1998PhysRevB}. This kind of peak is broad and appears as a hump in MHLs. Previous studies indicate that the second peak in YBCO corresponds to a transition of flux motion from elastic to plastic\cite{Abulafia1996PhysRevLett}. Additionally, several experiments revealed a kink of magnetization between the valley and the top of the second peak in MHL curves in YBCO\cite{Nishizaki1998PhysRevB,Nishizaki2000PhysRevB}, this kink was interpreted as an order-disorder transition just like the second peak found in Bi-2212. Furthermore, some experiments showed trace of multiple magnetization peaks in YBCO\cite{Deligiannis1997PhysRevLett,Sarkar2001PhysRevB}, TBCCO\cite{Pissas2006PhysRevB}, Hg-1201\cite{Stamopoulos2002PhysRevB} etc., which further complicates the explanation and understanding of the SP effect.\par

\begin{figure}
    \centering
    \setlength{\abovecaptionskip}{0cm}
    \setlength{\belowcaptionskip}{-0.2cm}
    \hspace{-0.7cm}
	\includegraphics[width=8cm]{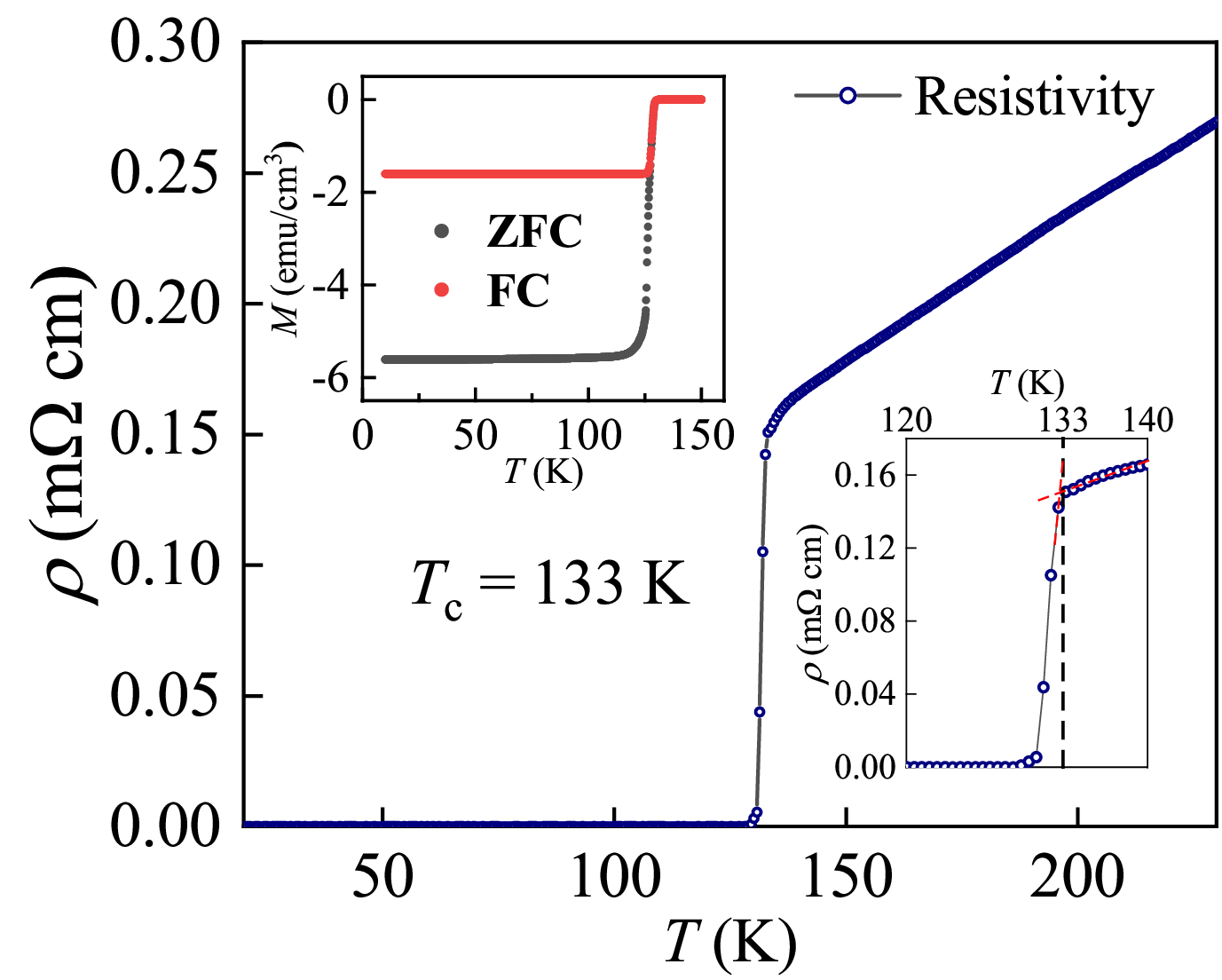}
	\caption{Main panel: Temperature dependence of the in-plain resistivity of a HgBa$_2$Ca$_2$Cu$_3$O$_8$ single crystal at zero magnetic field. The bottom-right inset gives an enlarged view of resistivity data near $T{\rm_c}$. The upper-left inset shows the temperature dependence of the magnetization of HgBa$_2$Ca$_2$Cu$_3$O$_8$ single crystal under a magnetic field of 10 Oe measured in zero-field-cooling (ZFC) and field-cooling (FC) modes. \label{fig1}}
	\end{figure}

\begin{figure*}
    \centering
    \hspace{-0.5cm}
    \setlength{\abovecaptionskip}{0.2cm}
    \setlength{\belowcaptionskip}{-0.1cm}
    \includegraphics[width=12cm]{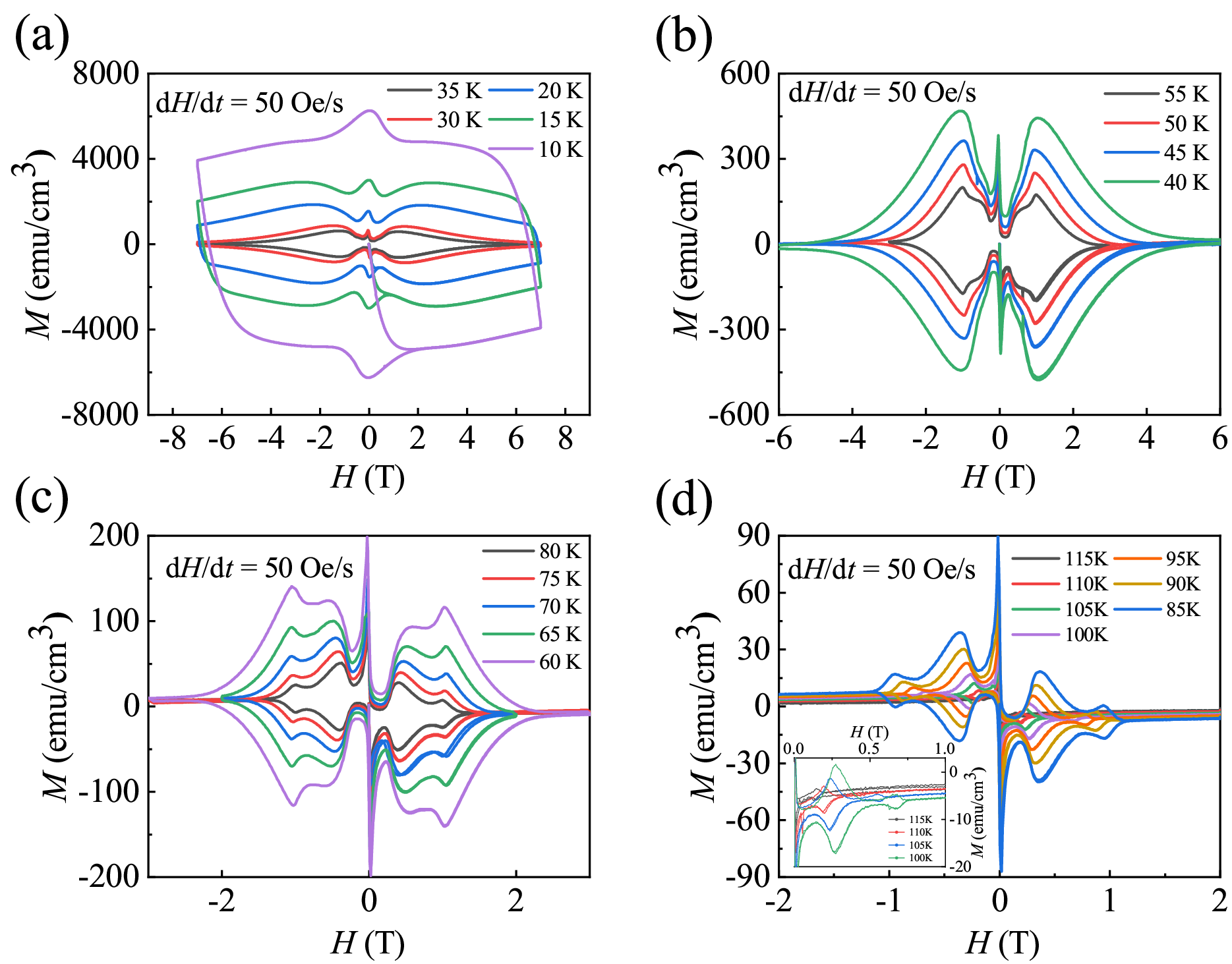}
    \caption{(a)$-$(d) Isothermal magnetic hysteresis loops of a HgBa$_2$Ca$_2$Cu$_3$O$_8$ single crystal for $H \parallel c$. Temperatures are indicated in each panel. The inset in (d) shows an enlarged view of magnetization in the temperature region from 100 K to 115 K where the third peak gets into a step-like structure and eventually vanishes at higher temperatures. All MHLs shown here were measured with ${\rm d} H/{\rm d}t = 50$ Oe/s.
} \label{fig2}
	\end{figure*}

In this Letter, we examine the vortex phase diagram of HgBa$_2$Ca$_2$Cu$_3$O$_8$ (Hg-1223) single crystals through detailed measurements of magnetization. For the first time, three magnetization peaks are clearly shown in isothermal MHLs in Hg-1223 single crystals. This three-peak effect appears in a wide temperature region with the highest temperature exceeding 100 K, the alternative evolution of the intensity of the second and third peaks reveals significant effects of thermal fluctuations on the flux motion. Our results indicate that the second peak is corresponding to the order-disorder transition from Bragg glass to vortex glass just like that in Bi-2212; while the third peak is associated with the elastic-plastic crossover, similar to the second peak in YBCO, which further evolves into a sharp transition due to thermal melting. Our results serve as a unified understanding concerning the SP effect in different systems.\par

Single-crystalline HgBa$_2$Ca$_2$Cu$_3$O$_8$ samples were synthesized in a high-pressure method using the same custom-built furnace mentioned in Refs.\cite{Wang2018PhysRevMaterials}. $T_{\rm c}$ of the as-grown samples is about 110K which can be tuned from 70K to 133K by post-annealing in air of oxygen atmosphere under different temperature condition. Optimized post-annealing conditions allowed our sample to exhibit both a sharp superconducting transition and the highest $T_{\rm c}$. Sample 1(S1) used for resistivity measurement and magnetization measurement at 10 Oe for both zero-field cooling (ZFC) and field-cooling (FC) has a mass of 0.41 mg with dimensions of 1.66$\times$0.64$\times$0.065 mm$^3$. Sample 2(S2) from the same batch used for MHLs measurement has a mass of 0.42 mg with dimensions of 1.3$\times$0.84$\times$0.065 mm$^3$. The magnetization was measured with a superconducting quantum interference device (SQUID-VSM, Quantum Design). The resistivity was measured with a Physical Property Measurement System (PPMS 16T, Quantum Design) by the standard four-probe method. 

The main panel of Fig. \ref{fig1} displays the resistivity $\rho$ as a function of temperature $T$ at zero field. The onset transition temperature taken with a criterion of $90\% \rho_{\rm n}$ at zero field is about 133 K, while the zero-resistance temperature is about 129 K. The enlarged view of resistivity near the superconducting transition is shown in the bottom-right inset of Fig. 1. The diamagnetic transition of the Hg-1223 single crystal is shown in the upper-left inset of Fig. \ref{fig1}. The ZFC curve shows perfect diamagnetism in the low temperature region and the sharp transition with the onset temperature $T_{\rm c}$ = 129 K. Both magnetic and resistive data indicate good quality of our samples.\par

\begin{figure*}
    \centering
    \hspace{-0.5cm}
    \setlength{\abovecaptionskip}{0.2cm}
    \setlength{\belowcaptionskip}{-0.1cm}
	\includegraphics[width=12cm]{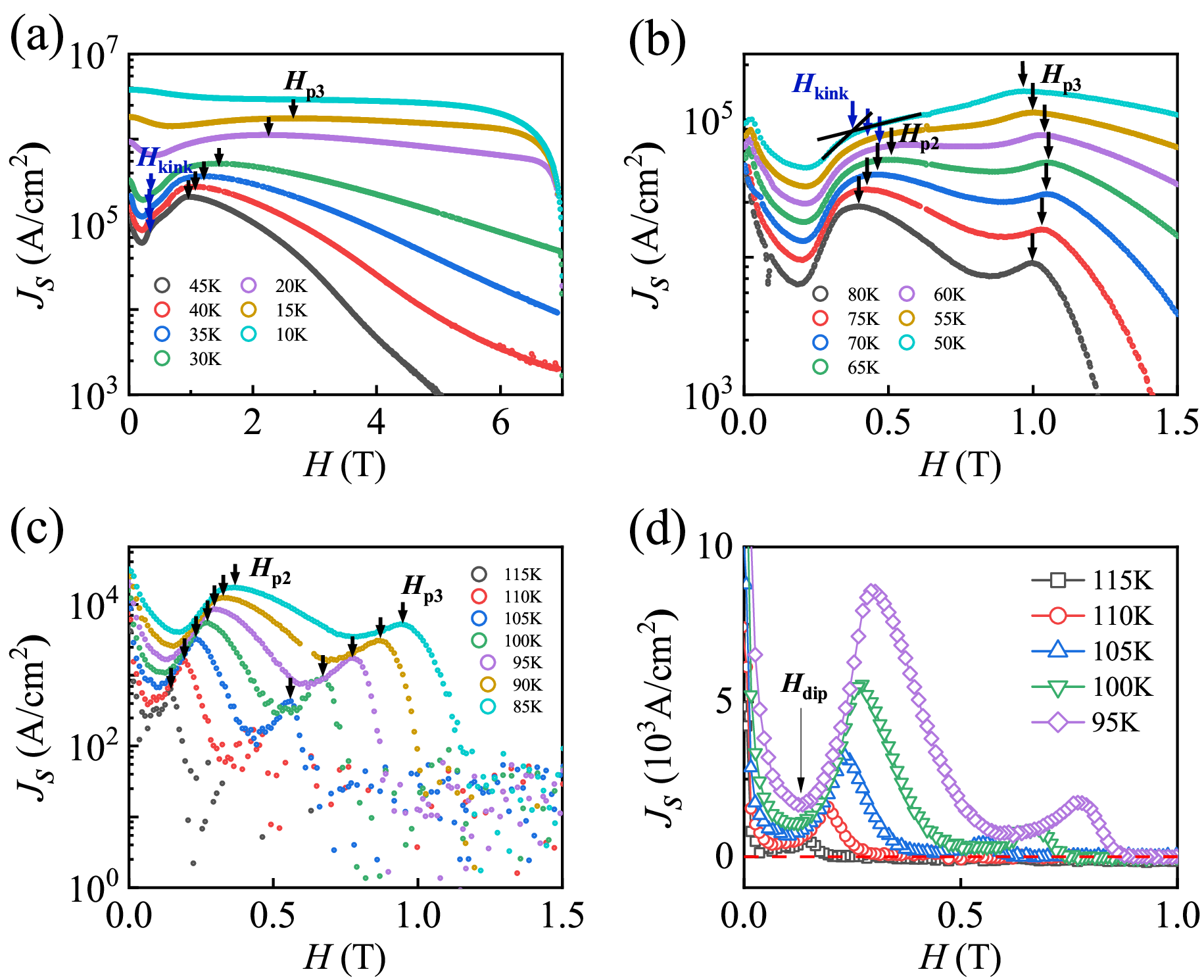}
	\caption{(a)$-$(d) Magnetic field dependence of transient superconducting current density $J_s$ derived from the Bean critical state model from Fig.\ref{fig2} in a semi-log plot for temperatures $T$ = 10$-$45 K(a), $T$ = 50$-$80 K(b), $T$ = 85$-$115 K(c). The arrows indicate positions of $H_{\rm kink}$, $H_{\rm p2}$, $H_{\rm p3}$, respectively. The panel (d) shows the $J_{\rm s}(H)$ curves at high temperatures using linear coordinates. Estimation of $H_{\rm kink}$ is show in panel (a) and (b).} \label{fig3}
	\end{figure*}

The isothermal MHLs for the magnetic field along c-axis from 10 K to 115 K with ${\rm d} H/{\rm d}t = 50$ Oe/s are displayed in Figs. \ref{fig2}(a)$–$(d). After cooling the sample to the required temperature at zero field, the measurements were made by increasing the magnetic field to $H_{\rm max}$ and then sweeping the field to $-H_{\rm max}$, finally the field was swept to $H_{\rm max}$ again to get the entire envelope of the MHL.  For $T<$ 60 K, the MHLs are symmetric about M = 0, indicating a strong bulk pinning; while for $T>$ 60 K, the MHLs gradually become asymmetric about zero magnetization, indicating that equilibrium magnetization of the surface layer plays an important role.

Based on the Bean critical state model\cite{BEAN1964RevModPhys}, we get the transient superconducting current density
 \begin{equation}
J_{\rm s} = 20\frac{\Delta M}{b(1-b/3a)}, 
     \end{equation}
     
     Where $\Delta M = M_{\rm down}-M_{\rm up}$ measured in emu/cm$^3$. $M_{\rm down}$ and $M_{\rm up}$ are the magnetization when sweeping fields down and up, respectively. $a$ and $b$ ($a > b$) represent the length and width of the sample measured in cm. Figs.~\ref{fig3}(a)$-$(c) present the field dependence of $J_{\rm s}$ at constant temperatures using semi-logarithmic coordinates. 

At all temperatures, the MHL exhibits a very sharp peak near zero field. This specific peak is noted as the first peak and is related to the strong pinning\cite{OvchinnikovPhysRevB,xing2020vortex}. Meanwhile, multiple magnetization peaks are clearly seen in MHLs for a wide temperature region (60 K to 105 K). As shown in Fig. \ref{fig2}(a) and (b), and Fig. \ref{fig3}(a) and (b), a kinky structure appears at temperatures between 30 K to 55 K. The field is marked by $H_{\rm kink}$ and is estimated as the crossover of two straight lines near the kink as shown in Fig. \ref{fig3}(b) for $J_{\rm s}-H$ at 50 K\cite{xing2020vortex}.
This kink gradually evolves into the second peak marked by $H_{\rm p2}$ at temperatures all the way up to 115 K. For clarity, the peak near zero field is called the first peak, the peak evolved from the kink at temperatures above about 60 K is called the second peak with a characteristic field $H_{\rm p2}$, and the peak at a higher magnetic field from 15 K to 110 K is called the third peak with a characteristic field $H_{\rm p3}$. In low temperature region ($T <$ 30K), only a huge and broad peak corresponding with $H_{\rm p3}$ exists, which is similar to the second peak in compounds HgBa$_2$CuO$_{4+\delta}$\cite{Cole2023PhysRevB} or YBCO~\cite{Nishizaki2000PhysRevB}.  With increasing temperature the magnitude of magnetization corresponding to the third peak shrinks, making the kinky structure transform to a clear peak at $H_{\rm p2}$. For 85 K $<T<$ 115 K, the second peak at $H_{\rm p2}$ remains visible while the third peak at $H_{\rm p3}$ evolves into a step-like structure and finally disappears at high temperatures. The inset in Figs. \ref{fig2} (d) shows the enlarged views of MHLs from 100 K to 115 K for $H>$ 0. Between $H_{\rm p2}$ and $H_{\rm p3}$, the narrow width of the MHL shows the similar reentrance behavior found in YBCO crystal\cite{Nishizaki1998PhysRevB}. Figs.~\ref{fig3}(d) shows $J_{\rm s}-H$ at high temperatures using linear coordinates. The horizontal dashed line representing $J_{\rm s}$ = 0 helps to locate $H_{\rm irr}$. The third peak gets into a step-like structure and gets close to $H_{\rm irr}$ just like the first order transition in YBCO\cite{Nishizaki2000PhysRevB}.

\begin{figure}
	\centerline{\includegraphics[width=7cm]{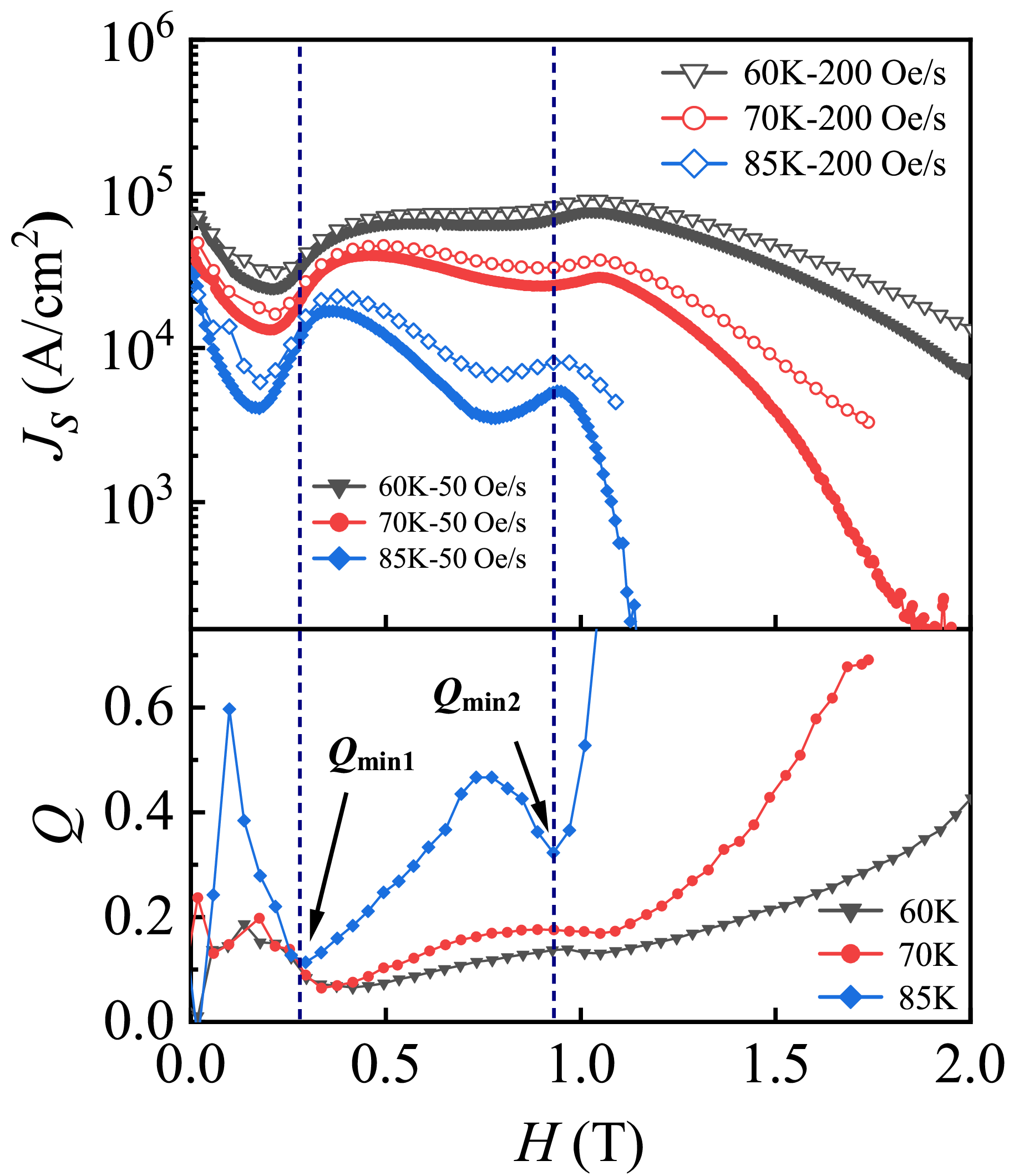}}
    \hspace{-0.5cm}
    \setlength{\abovecaptionskip}{-0.3cm}
    \setlength{\belowcaptionskip}{-0.1cm}
	\caption
 {Field dependence of transient superconducting current density $J_{\rm s}(H)$ with ${\rm d} H/{\rm d}t = $ 50 Oe/s (solid) and 200 Oe/s (open) and dynamic magnetic relaxation rate $Q(H)$ at 60K, 70K, and 85K. Two vertical dashed lines indicate the positions of $Q_{\rm min1}$ and $Q_{\rm min2}$ for 85K.
} \label{fig4}
	\end{figure}

For more in-depth analysis of the vortex dynamics, the dynamic relaxation rate $Q$ was measured. Different from the traditional magnetic relaxation rate $S=-{\rm d \thinspace ln}|M|/{\rm d \thinspace ln}t$, $Q$ is measured during the field sweeping process, which is defined as 
     \begin{equation}
Q \equiv \frac{{\rm d \thinspace ln} J}{{\rm d \thinspace ln(d}B/{\rm d}t)} = \frac{{\rm d \thinspace ln} \Delta M}{{\rm d \thinspace ln(d} B/{\rm d}t)}.
     \end{equation}\par
     
In the intermediate temperature region, the transient superconducting current density $J_{\rm s}(H)$ measured with two different field sweeping rates and dynamic magnetic relaxation rate $Q(H)$ are shown side by side in order to understand the underlying physics of the vortex transitions. As shown in Fig. \ref{fig4}, the first minimum $Q_{\rm min1}$ occurs just before the $H_{\rm p2}$ at a low field, strongly indicating a transition which suppresses the motion of vortex. In Bi-2212, the second peak $H_{\rm p2}$ is considered to be related to order-disorder transition, and the position of $H_{\rm p2}$ is almost independent of the temperature in the $H-T$ phase diagram\cite{Khaykovich1996VL,li2002bg}. In the clean YBCO single crystal, a kink that is related to the order-disorder transition was observed, which moves to higher field as the temperature increases\cite{Giller1999PhysRevB}. Similar behavior was also observed in other materials like La$_{2-x}$Sr$_x$CuO$_4$\cite{Radzyner2002PhysRevB} and Ba$_{1-x}$K$_x$Fe$_2$As$_2$\cite{Liu2024PhysRevB}. However, when the proper electron radiation was applied to the YBCO sample, the kink became relatively independent of $T$ in the low temperature region\cite{Nishizaki2000PhysRevB}. 
In the present case, it is reasonable to attribute ths minimum $Q_{\rm min1}$ to the increase of the $c$-axis vortex correlation length associating with the order-disorder transition\cite{Goffman1998PhysRevB}. 
Over the field for the second minimum $Q_{\rm min2}$, the dynamic magnetic relaxation rate increases at a fast speed indicating plastic vortex motion here. However, the dynamic magnetic relaxation rate increases gently between the fields for $Q_{\rm min1}$ and $Q_{\rm max2}$, which means that the vortex motion is still elastic.\par

\begin{figure}[h]
	\centerline{\includegraphics[width=8.3cm]{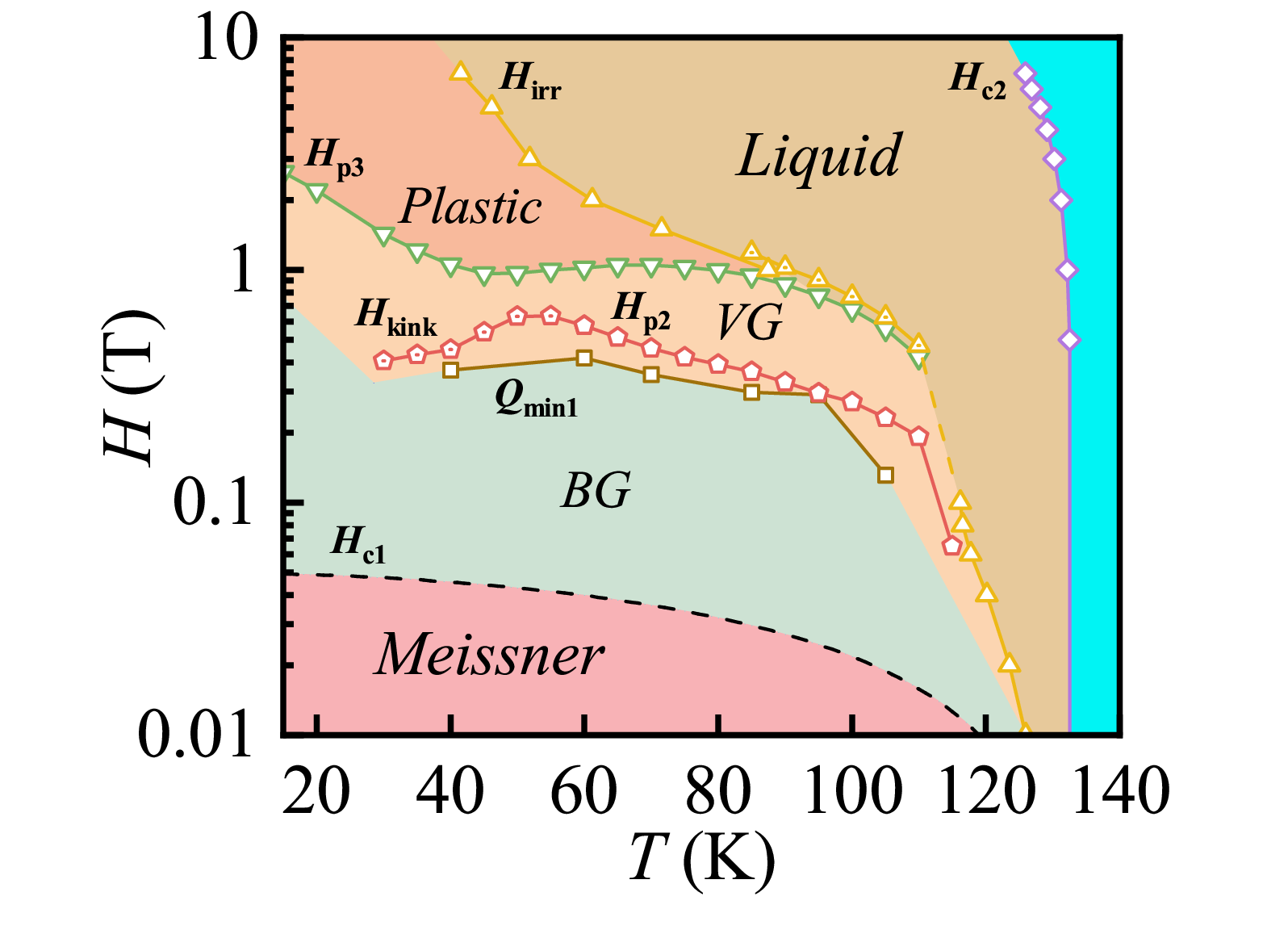}}
    \setlength{\abovecaptionskip}{0.1cm}
    \setlength{\belowcaptionskip}{-0.1cm}
	\caption
 { Vortex phase diagram of HgBa$_2$Ca$_2$Cu$_3$O$_8$ single crystal with $H \parallel c$ in a semi-log plot, depicted from the results of magnetization measurements and dynamic magnetic relaxation rate measurements. The transition lines $Q_{\rm min1}$(square) and $H_{\rm p3}$(down triangle) separate the vortex solid into three parts. ). 
} \label{fig5}
	\end{figure}

Finally, using the characteristic temperatures and fields obtained above, we construct the vortex phase diagram of Hg-1223 single crystal. The results are shown in Fig.~\ref{fig5}. Here the $H_{\rm c2}(T)$ (diamond) is determined by the deviating points of resistivity at fixed fields from the normal-state extrapolation line (not shown here). The irreversibility line $H_{\rm irr}$ is determined by the separation of $M \rm{_{ZFC}}$ and $M\rm{_{FC}}$ curves (up triangle) and also determined by taking a criterion of 10 A/cm$^2$ in $J_{\rm s}(H)$ curves for 85 K $< T <$ 115 K (up triangle with a point). The field values of $H_{\rm p2}$ (pentagon) and $H_{\rm p3}$ (down triangle) are located at the second peak and the third peak of $J_{\rm s}(H)$, respectively. The schematic curve of the lower critical field $H_{\rm c1}$ is represented by the black dashed line.

Now we give a general explanation to the occurrence of second and third magnetization peaks in wide temperature region. Below the field with $Q_{\rm min1}$, the vortex system may be quasi-ordered and called Bragg glass (BG)\cite{Giamarchi1994PhysRevLett,Giamarchi1995PhysRevB} which exhibits translational order despite the presence of some pinning centers. With increasing field, the interaction between vortices gets enhanced. Therefore, the vortex system should undergo a transition into the disordered vortex-glass (VG) phase, where the vortex matter can be described by the collective pinning theory. In the VG area, the vortex entities are pinned in a collective way\cite{Blatter1994,Koch1989PhysRevLett}, thus $J_{\rm s}$ is enhanced. In high temperature region, $J_{\rm s}$ quickly gets its maximum at $H_{\rm p2}$ and then goes down, while in the low temperature region, a kink instead of a peak appears. In this scenario, the kinky structure of magnetization in YBCO and our present sample reflects the vortex order-disorder transition. The difference between Hg-1223 and YBCO is that the kink in the present sample grows up as a second peak in high temperature region. Obviously, whether the order-disorder transition shows up as a kink or a peak strongly relies on the competition between the quenched disorder and the thermal fluctuations\cite{kwok2016vortices}. In low temperature region where the kink appears, the quenched disorders dominate so the entanglement is easily established leading to the increase of $J_{\rm s}(H)$ after the order-disorder transition, just like in YBCO. In the intermediate temperature region (e.g., 60 K $< T <$ 105 K), the kink has already evolved into a sharp peak at $H_{\rm p2}$ making the order-disorder transition more clear, which suggests that the vortex system may become more ordered with the help of thermal fluctuation. When the field is higher than $H_{\rm p2}$, the entanglement of vortices will enhance the critical current density, leading to the appearance of the third sharp peak at $H_{\rm p3}$. Further increase magnetic field will lead to the proliferation of the vortex dislocations and the vortex motion becomes plastic, accordingly the critical current density drops down quickly accompanied by the rapid increase of the dynamical relaxation rate. In the low temperature region, due to the lack of thermal fluctuation, this crossover occurs at much higher magnetic fields, thus the $H_{\rm p3}$(T) exhibits an upturn in low temperature region.In high temperature region, the strong thermal fluctuation will make the vortex dislocations proliferate significantly, leading to a suppressed third peak, and eventually this elastic-plastic crossover becomes a step-like thermal melting transition. The absence of the third peak in Bi-2212 may be attributed to its strongly layered structure and the entanglement is hard to be established after the second peak.

In conclusion, through the measurements of magnetization in a wide temperature range, we demonstrate for the first time that three magnetization peaks coexist in the MHLs in HgBa$_2$Ca$_2$Cu$_3$O$_8$ single crystal. Our data clearly indicate that the order-disorder transition occurs first as a kinky structure at low temperatures, and then it evolves to a second peak effect at high temperatures. However, the elastic-plastic transition occurs at higher magnetic fields due to the formation of the vortex dislocations. The evolution of the second peak from a kinky structure at low temperatures to a sharp peak at high temperatures could be explained by the thermal fluctuation effect. The ``ordered state'' at high temperatures may have a better translational order due to the thermal fluctuations. The shape of the third peak evolves also from a broad one at low temperatures to a sharp one at high temperatures, clearly showing the interplay of the quenched disorder and the thermal fluctuation effect. The observation of three magnetization peaks and the obtained phase diagram can unify the understanding of different "second peak" structures in different systems, supporting the universality of vortex phase diagram for HTSCs.
\section*{ACKNOWLEDGMENTS}
The project is supported by the National Natural Science Foundation of China (Nos. 1243000380, 12434004, 11927809, 12061131004, 11888101), National Key Research and Development Program of China  (No. 2022YFA1403201,2021YFA1401900). W. H. acknowledges support from the Postdoctoral Innovative Talent program (BX2021018) and the China Postdoctoral Science Foundation (2021M700250).

\bibliography{reference}

\begin{thebibliography}{41}%
\makeatletter
\providecommand \@ifxundefined [1]{%
 \@ifx{#1\undefined}
}%
\providecommand \@ifnum [1]{%
 \ifnum #1\expandafter \@firstoftwo
 \else \expandafter \@secondoftwo
 \fi
}%
\providecommand \@ifx [1]{%
 \ifx #1\expandafter \@firstoftwo
 \else \expandafter \@secondoftwo
 \fi
}%
\providecommand \natexlab [1]{#1}%
\providecommand \enquote  [1]{``#1''}%
\providecommand \bibnamefont  [1]{#1}%
\providecommand \bibfnamefont [1]{#1}%
\providecommand \citenamefont [1]{#1}%
\providecommand \href@noop [0]{\@secondoftwo}%
\providecommand \href [0]{\begingroup \@sanitize@url \@href}%
\providecommand \@href[1]{\@@startlink{#1}\@@href}%
\providecommand \@@href[1]{\endgroup#1\@@endlink}%
\providecommand \@sanitize@url [0]{\catcode `\\12\catcode `\$12\catcode
  `\&12\catcode `\#12\catcode `\^12\catcode `\_12\catcode `\%12\relax}%
\providecommand \@@startlink[1]{}%
\providecommand \@@endlink[0]{}%
\providecommand \url  [0]{\begingroup\@sanitize@url \@url }%
\providecommand \@url [1]{\endgroup\@href {#1}{\urlprefix }}%
\providecommand \urlprefix  [0]{URL }%
\providecommand \Eprint [0]{\href }%
\providecommand \doibase [0]{https://doi.org/}%
\providecommand \selectlanguage [0]{\@gobble}%
\providecommand \bibinfo  [0]{\@secondoftwo}%
\providecommand \bibfield  [0]{\@secondoftwo}%
\providecommand \translation [1]{[#1]}%
\providecommand \BibitemOpen [0]{}%
\providecommand \bibitemStop [0]{}%
\providecommand \bibitemNoStop [0]{.\EOS\space}%
\providecommand \EOS [0]{\spacefactor3000\relax}%
\providecommand \BibitemShut  [1]{\csname bibitem#1\endcsname}%
\let\auto@bib@innerbib\@empty
\bibitem [{\citenamefont {Blatter}\ \emph {et~al.}(1994)\citenamefont
  {Blatter}, \citenamefont {Feigel'man}, \citenamefont {Geshkenbein},
  \citenamefont {Larkin},\ and\ \citenamefont {Vinokur}}]{Blatter1994}%
  \BibitemOpen
  \bibfield  {author} {\bibinfo {author} {\bibfnamefont {G.}~\bibnamefont
  {Blatter}}, \bibinfo {author} {\bibfnamefont {M.~V.}\ \bibnamefont
  {Feigel'man}}, \bibinfo {author} {\bibfnamefont {V.~B.}\ \bibnamefont
  {Geshkenbein}}, \bibinfo {author} {\bibfnamefont {A.~I.}\ \bibnamefont
  {Larkin}},\ and\ \bibinfo {author} {\bibfnamefont {V.~M.}\ \bibnamefont
  {Vinokur}},\ }\bibfield  {title} {\bibinfo {title} {{Vortices in
  high-temperature superconductors}},\ }\href
  {https://doi.org/10.1103/RevModPhys.66.1125} {\bibfield  {journal} {\bibinfo
  {journal} {Rev. Mod. Phys.}\ }\textbf {\bibinfo {volume} {66}},\ \bibinfo
  {pages} {1125} (\bibinfo {year} {1994})}\BibitemShut {NoStop}%
\bibitem [{\citenamefont {Vinokur}\ \emph {et~al.}(1998)\citenamefont
  {Vinokur}, \citenamefont {Khaykovich}, \citenamefont {Zeldov}, \citenamefont
  {Konczykowski}, \citenamefont {Doyle},\ and\ \citenamefont
  {Kes}}]{VINOKUR1998}%
  \BibitemOpen
  \bibfield  {author} {\bibinfo {author} {\bibfnamefont {V.}~\bibnamefont
  {Vinokur}}, \bibinfo {author} {\bibfnamefont {B.}~\bibnamefont {Khaykovich}},
  \bibinfo {author} {\bibfnamefont {E.}~\bibnamefont {Zeldov}}, \bibinfo
  {author} {\bibfnamefont {M.}~\bibnamefont {Konczykowski}}, \bibinfo {author}
  {\bibfnamefont {R.~A.}\ \bibnamefont {Doyle}},\ and\ \bibinfo {author}
  {\bibfnamefont {P.~H.}\ \bibnamefont {Kes}},\ }\bibfield  {title} {\bibinfo
  {title} {{Lindemann criterion and vortex-matter phase transitions in
  high-temperature superconductors}},\ }\href
  {https://doi.org/https://doi.org/10.1016/S0921-4534(97)01795-4} {\bibfield
  {journal} {\bibinfo  {journal} {Physica C: Superconductivity}\ }\textbf
  {\bibinfo {volume} {295}},\ \bibinfo {pages} {209} (\bibinfo {year}
  {1998})}\BibitemShut {NoStop}%
\bibitem [{\citenamefont {Klein}\ \emph {et~al.}(2001)\citenamefont {Klein},
  \citenamefont {Joumard}, \citenamefont {Blanchard}, \citenamefont {Marcus},
  \citenamefont {Cubitt}, \citenamefont {Giamarchi},\ and\ \citenamefont
  {Le~Doussal}}]{klein2001bragg}%
  \BibitemOpen
  \bibfield  {author} {\bibinfo {author} {\bibfnamefont {T.}~\bibnamefont
  {Klein}}, \bibinfo {author} {\bibfnamefont {I.}~\bibnamefont {Joumard}},
  \bibinfo {author} {\bibfnamefont {S.}~\bibnamefont {Blanchard}}, \bibinfo
  {author} {\bibfnamefont {J.}~\bibnamefont {Marcus}}, \bibinfo {author}
  {\bibfnamefont {R.}~\bibnamefont {Cubitt}}, \bibinfo {author} {\bibfnamefont
  {T.}~\bibnamefont {Giamarchi}},\ and\ \bibinfo {author} {\bibfnamefont
  {P.}~\bibnamefont {Le~Doussal}},\ }\bibfield  {title} {\bibinfo {title} {{A
  Bragg glass phase in the vortex lattice of a type II superconductor}},\
  }\href {https://www.nature.com/articles/35096534} {\bibfield  {journal}
  {\bibinfo  {journal} {Nature}\ }\textbf {\bibinfo {volume} {413}},\ \bibinfo
  {pages} {404} (\bibinfo {year} {2001})}\BibitemShut {NoStop}%
\bibitem [{\citenamefont {Avraham}\ \emph {et~al.}(2001)\citenamefont
  {Avraham}, \citenamefont {Khaykovich}, \citenamefont {Myasoedov},
  \citenamefont {Rappaport}, \citenamefont {Shtrikman}, \citenamefont
  {Feldman}, \citenamefont {Tamegai}, \citenamefont {Kes}, \citenamefont {Li},
  \citenamefont {Konczykowski}, \citenamefont {van~der Beek},\ and\
  \citenamefont {Zeldov}}]{avraham2001inverse}%
  \BibitemOpen
  \bibfield  {author} {\bibinfo {author} {\bibfnamefont {N.}~\bibnamefont
  {Avraham}}, \bibinfo {author} {\bibfnamefont {B.}~\bibnamefont {Khaykovich}},
  \bibinfo {author} {\bibfnamefont {Y.}~\bibnamefont {Myasoedov}}, \bibinfo
  {author} {\bibfnamefont {M.}~\bibnamefont {Rappaport}}, \bibinfo {author}
  {\bibfnamefont {H.}~\bibnamefont {Shtrikman}}, \bibinfo {author}
  {\bibfnamefont {D.~E.}\ \bibnamefont {Feldman}}, \bibinfo {author}
  {\bibfnamefont {T.}~\bibnamefont {Tamegai}}, \bibinfo {author} {\bibfnamefont
  {P.~H.}\ \bibnamefont {Kes}}, \bibinfo {author} {\bibfnamefont
  {M.}~\bibnamefont {Li}}, \bibinfo {author} {\bibfnamefont {M.}~\bibnamefont
  {Konczykowski}}, \bibinfo {author} {\bibfnamefont {K.}~\bibnamefont {van~der
  Beek}},\ and\ \bibinfo {author} {\bibfnamefont {E.}~\bibnamefont {Zeldov}},\
  }\bibfield  {title} {\bibinfo {title} {{`Inverse' melting of a vortex
  lattice}},\ }\href {https://www.nature.com/articles/35078021} {\bibfield
  {journal} {\bibinfo  {journal} {Nature}\ }\textbf {\bibinfo {volume} {411}},\
  \bibinfo {pages} {451} (\bibinfo {year} {2001})}\BibitemShut {NoStop}%
\bibitem [{\citenamefont {Beidenkopf}\ \emph {et~al.}(2005)\citenamefont
  {Beidenkopf}, \citenamefont {Avraham}, \citenamefont {Myasoedov},
  \citenamefont {Shtrikman}, \citenamefont {Zeldov}, \citenamefont
  {Rosenstein}, \citenamefont {Brandt},\ and\ \citenamefont
  {Tamegai}}]{beidenkopf2005equilibrium}%
  \BibitemOpen
  \bibfield  {author} {\bibinfo {author} {\bibfnamefont {H.}~\bibnamefont
  {Beidenkopf}}, \bibinfo {author} {\bibfnamefont {N.}~\bibnamefont {Avraham}},
  \bibinfo {author} {\bibfnamefont {Y.}~\bibnamefont {Myasoedov}}, \bibinfo
  {author} {\bibfnamefont {H.}~\bibnamefont {Shtrikman}}, \bibinfo {author}
  {\bibfnamefont {E.}~\bibnamefont {Zeldov}}, \bibinfo {author} {\bibfnamefont
  {B.}~\bibnamefont {Rosenstein}}, \bibinfo {author} {\bibfnamefont {E.~H.}\
  \bibnamefont {Brandt}},\ and\ \bibinfo {author} {\bibfnamefont
  {T.}~\bibnamefont {Tamegai}},\ }\bibfield  {title} {\bibinfo {title}
  {{Equilibrium First-Order Melting and Second-Order Glass Transitions of the
  Vortex Matter in
  ${\mathrm{Bi}}_{2}{\mathrm{Sr}}_{2}{\mathrm{CaCu}}_{2}{\mathrm{O}}_{8}$}},\
  }\href {https://doi.org/10.1103/PhysRevLett.95.257004} {\bibfield  {journal}
  {\bibinfo  {journal} {Phys. Rev. Lett.}\ }\textbf {\bibinfo {volume} {95}},\
  \bibinfo {pages} {257004} (\bibinfo {year} {2005})}\BibitemShut {NoStop}%
\bibitem [{\citenamefont {LeBlanc}\ and\ \citenamefont
  {Little}(1961)}]{LeBlancandLittle1961}%
  \BibitemOpen
  \bibfield  {author} {\bibinfo {author} {\bibfnamefont {M.~A.~R.}\
  \bibnamefont {LeBlanc}}\ and\ \bibinfo {author} {\bibfnamefont {W.~A.}\
  \bibnamefont {Little}},\ }\href@noop {} {\emph {\bibinfo {title} {Proceedings
  of the Seventh International Conference on Low-Temperature Physics, 1960,
  edited by G. M. Graham and A. C. Hollis, p.198}}}\ (\bibinfo  {publisher}
  {University of Toronto Press},\ \bibinfo {year} {1961})\BibitemShut {NoStop}%
\bibitem [{\citenamefont {Kopylov}\ \emph {et~al.}(1990)\citenamefont
  {Kopylov}, \citenamefont {Koshelev}, \citenamefont {Schegolev},\ and\
  \citenamefont {Togonidze}}]{KOPYLOV1990291}%
  \BibitemOpen
  \bibfield  {author} {\bibinfo {author} {\bibfnamefont {V.~N.}\ \bibnamefont
  {Kopylov}}, \bibinfo {author} {\bibfnamefont {A.~E.}\ \bibnamefont
  {Koshelev}}, \bibinfo {author} {\bibfnamefont {I.~F.}\ \bibnamefont
  {Schegolev}},\ and\ \bibinfo {author} {\bibfnamefont {T.~G.}\ \bibnamefont
  {Togonidze}},\ }\bibfield  {title} {\bibinfo {title} {{The role of surface
  effects in magnetization of high-$T_{\mathrm{c}}$ superconductors}},\ }\href
  {https://doi.org/https://doi.org/10.1016/0921-4534(90)90326-A} {\bibfield
  {journal} {\bibinfo  {journal} {Physica C: Superconductivity}\ }\textbf
  {\bibinfo {volume} {170}},\ \bibinfo {pages} {291} (\bibinfo {year}
  {1990})}\BibitemShut {NoStop}%
\bibitem [{\citenamefont {Kokkaliaris}\ \emph {et~al.}(2000)\citenamefont
  {Kokkaliaris}, \citenamefont {Zhukov}, \citenamefont {de~Groot},
  \citenamefont {Gagnon}, \citenamefont {Taillefer},\ and\ \citenamefont
  {Wolf}}]{kokkaliaris2000transition}%
  \BibitemOpen
  \bibfield  {author} {\bibinfo {author} {\bibfnamefont {S.}~\bibnamefont
  {Kokkaliaris}}, \bibinfo {author} {\bibfnamefont {A.~A.}\ \bibnamefont
  {Zhukov}}, \bibinfo {author} {\bibfnamefont {P.~A.~J.}\ \bibnamefont
  {de~Groot}}, \bibinfo {author} {\bibfnamefont {R.}~\bibnamefont {Gagnon}},
  \bibinfo {author} {\bibfnamefont {L.}~\bibnamefont {Taillefer}},\ and\
  \bibinfo {author} {\bibfnamefont {T.}~\bibnamefont {Wolf}},\ }\bibfield
  {title} {\bibinfo {title} {{Transition from elastic to plastic vortex phase
  and its evolution with quenched disorder in
  ${\mathrm{YBa}}_{2}{\mathrm{Cu}}_{3}{\mathrm{O}}_{y}$ single crystals}},\
  }\href {https://doi.org/10.1103/PhysRevB.61.3655} {\bibfield  {journal}
  {\bibinfo  {journal} {Phys. Rev. B}\ }\textbf {\bibinfo {volume} {61}},\
  \bibinfo {pages} {3655} (\bibinfo {year} {2000})}\BibitemShut {NoStop}%
\bibitem [{\citenamefont {Banerjee}\ \emph {et~al.}(2000)\citenamefont
  {Banerjee}, \citenamefont {Ramakrishnan}, \citenamefont {Grover},
  \citenamefont {Ravikumar}, \citenamefont {Mishra}, \citenamefont {Sahni},
  \citenamefont {Tomy}, \citenamefont {Balakrishnan}, \citenamefont {Paul},
  \citenamefont {Gammel}, \citenamefont {Bishop}, \citenamefont {Bucher},
  \citenamefont {Higgins},\ and\ \citenamefont
  {Bhattacharya}}]{Banerjee2000Peak}%
  \BibitemOpen
  \bibfield  {author} {\bibinfo {author} {\bibfnamefont {S.~S.}\ \bibnamefont
  {Banerjee}}, \bibinfo {author} {\bibfnamefont {S.}~\bibnamefont
  {Ramakrishnan}}, \bibinfo {author} {\bibfnamefont {A.~K.}\ \bibnamefont
  {Grover}}, \bibinfo {author} {\bibfnamefont {G.}~\bibnamefont {Ravikumar}},
  \bibinfo {author} {\bibfnamefont {P.~K.}\ \bibnamefont {Mishra}}, \bibinfo
  {author} {\bibfnamefont {V.~C.}\ \bibnamefont {Sahni}}, \bibinfo {author}
  {\bibfnamefont {C.~V.}\ \bibnamefont {Tomy}}, \bibinfo {author}
  {\bibfnamefont {G.}~\bibnamefont {Balakrishnan}}, \bibinfo {author}
  {\bibfnamefont {D.~M.}\ \bibnamefont {Paul}}, \bibinfo {author}
  {\bibfnamefont {P.~L.}\ \bibnamefont {Gammel}}, \bibinfo {author}
  {\bibfnamefont {D.~J.}\ \bibnamefont {Bishop}}, \bibinfo {author}
  {\bibfnamefont {E.}~\bibnamefont {Bucher}}, \bibinfo {author} {\bibfnamefont
  {M.~J.}\ \bibnamefont {Higgins}},\ and\ \bibinfo {author} {\bibfnamefont
  {S.}~\bibnamefont {Bhattacharya}},\ }\bibfield  {title} {\bibinfo {title}
  {{Peak effect, plateau effect, and fishtail anomaly: The reentrant
  amorphization of vortex matter in $2H\ensuremath{-}{\mathrm{NbSe}}_{2}$}},\
  }\href {https://doi.org/10.1103/PhysRevB.62.11838} {\bibfield  {journal}
  {\bibinfo  {journal} {Phys. Rev. B}\ }\textbf {\bibinfo {volume} {62}},\
  \bibinfo {pages} {11838} (\bibinfo {year} {2000})}\BibitemShut {NoStop}%
\bibitem [{\citenamefont {Yang}\ \emph {et~al.}(2008)\citenamefont {Yang},
  \citenamefont {Luo}, \citenamefont {Wang},\ and\ \citenamefont
  {Wen}}]{Yang2008PE}%
  \BibitemOpen
  \bibfield  {author} {\bibinfo {author} {\bibfnamefont {H.}~\bibnamefont
  {Yang}}, \bibinfo {author} {\bibfnamefont {H.}~\bibnamefont {Luo}}, \bibinfo
  {author} {\bibfnamefont {Z.}~\bibnamefont {Wang}},\ and\ \bibinfo {author}
  {\bibfnamefont {H.-H.}\ \bibnamefont {Wen}},\ }\bibfield  {title} {\bibinfo
  {title} {{Fishtail effect and the vortex phase diagram of single crystal
  ${\mathrm{Ba}}_{0.6}{\mathrm{K}}_{0.4}{\mathrm{Fe}}_{2}{\mathrm{As}}_{2}$
  }},\ }\href {https://doi.org/10.1063/1.2996576} {\bibfield  {journal}
  {\bibinfo  {journal} {Appl. Phys. Lett.}\ }\textbf {\bibinfo {volume} {93}},\
  \bibinfo {pages} {142506} (\bibinfo {year} {2008})}\BibitemShut {NoStop}%
\bibitem [{\citenamefont {Bonura}\ \emph {et~al.}(2012)\citenamefont {Bonura},
  \citenamefont {Giannini}, \citenamefont {Viennois},\ and\ \citenamefont
  {Senatore}}]{Bonura2012PE}%
  \BibitemOpen
  \bibfield  {author} {\bibinfo {author} {\bibfnamefont {M.}~\bibnamefont
  {Bonura}}, \bibinfo {author} {\bibfnamefont {E.}~\bibnamefont {Giannini}},
  \bibinfo {author} {\bibfnamefont {R.}~\bibnamefont {Viennois}},\ and\
  \bibinfo {author} {\bibfnamefont {C.}~\bibnamefont {Senatore}},\ }\bibfield
  {title} {\bibinfo {title} {{Temperature and time scaling of the peak-effect
  vortex configuration in FeTe$_{0.7}$Se$_{0.3}$}},\ }\href
  {https://doi.org/10.1103/PhysRevB.85.134532} {\bibfield  {journal} {\bibinfo
  {journal} {Phys. Rev. B}\ }\textbf {\bibinfo {volume} {85}},\ \bibinfo
  {pages} {134532} (\bibinfo {year} {2012})}\BibitemShut {NoStop}%
\bibitem [{\citenamefont {Chowdhury}\ \emph {et~al.}(2003)\citenamefont
  {Chowdhury}, \citenamefont {Kim}, \citenamefont {Jo},\ and\ \citenamefont
  {Lee}}]{chowdhury2003peak}%
  \BibitemOpen
  \bibfield  {author} {\bibinfo {author} {\bibfnamefont {P.}~\bibnamefont
  {Chowdhury}}, \bibinfo {author} {\bibfnamefont {H.-J.}\ \bibnamefont {Kim}},
  \bibinfo {author} {\bibfnamefont {I.-S.}\ \bibnamefont {Jo}},\ and\ \bibinfo
  {author} {\bibfnamefont {S.-I.}\ \bibnamefont {Lee}},\ }\bibfield  {title}
  {\bibinfo {title} {{Peak anomaly and irreversible magnetization in
  Tl$_{2}$Ba$_{2}$CaCu$_{2}$O$_{8}$ single crystals}},\ }\href
  {https://www.sciencedirect.com/science/article/pii/S0921453402019974}
  {\bibfield  {journal} {\bibinfo  {journal} {Physica C: Superconductivity}\
  }\textbf {\bibinfo {volume} {384}},\ \bibinfo {pages} {411} (\bibinfo {year}
  {2003})}\BibitemShut {NoStop}%
\bibitem [{\citenamefont {Ghosh}\ \emph {et~al.}(1996)\citenamefont {Ghosh},
  \citenamefont {Ramakrishnan}, \citenamefont {Grover}, \citenamefont {Menon},
  \citenamefont {Chandra}, \citenamefont {Chandrasekhar~Rao}, \citenamefont
  {Ravikumar}, \citenamefont {Mishra}, \citenamefont {Sahni}, \citenamefont
  {Tomy}, \citenamefont {Balakrishnan}, \citenamefont {Mck~Paul},\ and\
  \citenamefont {Bhattacharya}}]{Ghosh1996peak}%
  \BibitemOpen
  \bibfield  {author} {\bibinfo {author} {\bibfnamefont {K.}~\bibnamefont
  {Ghosh}}, \bibinfo {author} {\bibfnamefont {S.}~\bibnamefont {Ramakrishnan}},
  \bibinfo {author} {\bibfnamefont {A.~K.}\ \bibnamefont {Grover}}, \bibinfo
  {author} {\bibfnamefont {G.~I.}\ \bibnamefont {Menon}}, \bibinfo {author}
  {\bibfnamefont {G.}~\bibnamefont {Chandra}}, \bibinfo {author} {\bibfnamefont
  {T.~V.}\ \bibnamefont {Chandrasekhar~Rao}}, \bibinfo {author} {\bibfnamefont
  {G.}~\bibnamefont {Ravikumar}}, \bibinfo {author} {\bibfnamefont {P.~K.}\
  \bibnamefont {Mishra}}, \bibinfo {author} {\bibfnamefont {V.~C.}\
  \bibnamefont {Sahni}}, \bibinfo {author} {\bibfnamefont {C.~V.}\ \bibnamefont
  {Tomy}}, \bibinfo {author} {\bibfnamefont {G.}~\bibnamefont {Balakrishnan}},
  \bibinfo {author} {\bibfnamefont {D.}~\bibnamefont {Mck~Paul}},\ and\
  \bibinfo {author} {\bibfnamefont {S.}~\bibnamefont {Bhattacharya}},\
  }\bibfield  {title} {\bibinfo {title} {{Reentrant Peak Effect and Melting of
  a Flux Line Lattice in 2H-Nb ${\mathrm{Se}}_{2}$}},\ }\href
  {https://doi.org/10.1103/PhysRevLett.76.4600} {\bibfield  {journal} {\bibinfo
   {journal} {Phys. Rev. Lett.}\ }\textbf {\bibinfo {volume} {76}},\ \bibinfo
  {pages} {4600} (\bibinfo {year} {1996})}\BibitemShut {NoStop}%
\bibitem [{\citenamefont {Banerjee}\ \emph {et~al.}(1998)\citenamefont
  {Banerjee}, \citenamefont {Patil}, \citenamefont {Saha}, \citenamefont
  {Ramakrishnan}, \citenamefont {Grover}, \citenamefont {Bhattacharya},
  \citenamefont {Ravikumar}, \citenamefont {Mishra}, \citenamefont
  {Chandrasekhar~Rao}, \citenamefont {Sahni}, \citenamefont {Higgins},
  \citenamefont {Yamamoto}, \citenamefont {Haga}, \citenamefont {Hedo},
  \citenamefont {Inada},\ and\ \citenamefont {Onuki}}]{Banerjee1998peak}%
  \BibitemOpen
  \bibfield  {author} {\bibinfo {author} {\bibfnamefont {S.~S.}\ \bibnamefont
  {Banerjee}}, \bibinfo {author} {\bibfnamefont {N.~G.}\ \bibnamefont {Patil}},
  \bibinfo {author} {\bibfnamefont {S.}~\bibnamefont {Saha}}, \bibinfo {author}
  {\bibfnamefont {S.}~\bibnamefont {Ramakrishnan}}, \bibinfo {author}
  {\bibfnamefont {A.~K.}\ \bibnamefont {Grover}}, \bibinfo {author}
  {\bibfnamefont {S.}~\bibnamefont {Bhattacharya}}, \bibinfo {author}
  {\bibfnamefont {G.}~\bibnamefont {Ravikumar}}, \bibinfo {author}
  {\bibfnamefont {P.~K.}\ \bibnamefont {Mishra}}, \bibinfo {author}
  {\bibfnamefont {T.~V.}\ \bibnamefont {Chandrasekhar~Rao}}, \bibinfo {author}
  {\bibfnamefont {V.~C.}\ \bibnamefont {Sahni}}, \bibinfo {author}
  {\bibfnamefont {M.~J.}\ \bibnamefont {Higgins}}, \bibinfo {author}
  {\bibfnamefont {E.}~\bibnamefont {Yamamoto}}, \bibinfo {author}
  {\bibfnamefont {Y.}~\bibnamefont {Haga}}, \bibinfo {author} {\bibfnamefont
  {M.}~\bibnamefont {Hedo}}, \bibinfo {author} {\bibfnamefont {Y.}~\bibnamefont
  {Inada}},\ and\ \bibinfo {author} {\bibfnamefont {Y.}~\bibnamefont {Onuki}},\
  }\bibfield  {title} {\bibinfo {title} {{Anomalous peak effect in
  ${\mathrm{CeRu}}_{2}$ and $2H\ensuremath{-}{\mathrm{NbSe}}_{2}:$ Fracturing
  of a flux line lattice}},\ }\href {https://doi.org/10.1103/PhysRevB.58.995}
  {\bibfield  {journal} {\bibinfo  {journal} {Phys. Rev. B}\ }\textbf {\bibinfo
  {volume} {58}},\ \bibinfo {pages} {995} (\bibinfo {year} {1998})}\BibitemShut
  {NoStop}%
\bibitem [{\citenamefont {Bhattacharya}\ and\ \citenamefont
  {Higgins}(1993)}]{Bhattacharya1993disorder}%
  \BibitemOpen
  \bibfield  {author} {\bibinfo {author} {\bibfnamefont {S.}~\bibnamefont
  {Bhattacharya}}\ and\ \bibinfo {author} {\bibfnamefont {M.~J.}\ \bibnamefont
  {Higgins}},\ }\bibfield  {title} {\bibinfo {title} {Dynamics of a disordered
  flux line lattice},\ }\href {https://doi.org/10.1103/PhysRevLett.70.2617}
  {\bibfield  {journal} {\bibinfo  {journal} {Phys. Rev. Lett.}\ }\textbf
  {\bibinfo {volume} {70}},\ \bibinfo {pages} {2617} (\bibinfo {year}
  {1993})}\BibitemShut {NoStop}%
\bibitem [{\citenamefont {Troyanovski}\ \emph {et~al.}(2002)\citenamefont
  {Troyanovski}, \citenamefont {van Hecke}, \citenamefont {Saha}, \citenamefont
  {Aarts},\ and\ \citenamefont {Kes}}]{Troyanovski2002STM}%
  \BibitemOpen
  \bibfield  {author} {\bibinfo {author} {\bibfnamefont {A.~M.}\ \bibnamefont
  {Troyanovski}}, \bibinfo {author} {\bibfnamefont {M.}~\bibnamefont {van
  Hecke}}, \bibinfo {author} {\bibfnamefont {N.}~\bibnamefont {Saha}}, \bibinfo
  {author} {\bibfnamefont {J.}~\bibnamefont {Aarts}},\ and\ \bibinfo {author}
  {\bibfnamefont {P.~H.}\ \bibnamefont {Kes}},\ }\bibfield  {title} {\bibinfo
  {title} {{STM Imaging of Flux Line Arrangements in the Peak Effect Regime}},\
  }\href {https://doi.org/10.1103/PhysRevLett.89.147006} {\bibfield  {journal}
  {\bibinfo  {journal} {Phys. Rev. Lett.}\ }\textbf {\bibinfo {volume} {89}},\
  \bibinfo {pages} {147006} (\bibinfo {year} {2002})}\BibitemShut {NoStop}%
\bibitem [{\citenamefont {Pippard}(1969)}]{pippard1969PE}%
  \BibitemOpen
  \bibfield  {author} {\bibinfo {author} {\bibfnamefont {A.~B.}\ \bibnamefont
  {Pippard}},\ }\bibfield  {title} {\bibinfo {title} {{A possible mechanism for
  the peak effect in type II superconductors}},\ }\href
  {https://www.tandfonline.com/doi/abs/10.1080/14786436908217779} {\bibfield
  {journal} {\bibinfo  {journal} {Philos. Mag.}\ }\textbf {\bibinfo {volume}
  {19}},\ \bibinfo {pages} {217} (\bibinfo {year} {1969})}\BibitemShut
  {NoStop}%
\bibitem [{\citenamefont {Larkin}\ and\ \citenamefont
  {Ovchinnikov}(1979)}]{larkin1979pinning}%
  \BibitemOpen
  \bibfield  {author} {\bibinfo {author} {\bibfnamefont {A.~I.}\ \bibnamefont
  {Larkin}}\ and\ \bibinfo {author} {\bibfnamefont {Y.~N.}\ \bibnamefont
  {Ovchinnikov}},\ }\bibfield  {title} {\bibinfo {title} {{Pinning in type II
  superconductors}},\ }\href
  {https://link.springer.com/article/10.1007/BF00117160} {\bibfield  {journal}
  {\bibinfo  {journal} {J. Low Temp. Phys.}\ }\textbf {\bibinfo {volume}
  {34}},\ \bibinfo {pages} {409} (\bibinfo {year} {1979})}\BibitemShut
  {NoStop}%
\bibitem [{\citenamefont {Khaykovich}\ \emph {et~al.}(1996)\citenamefont
  {Khaykovich}, \citenamefont {Zeldov}, \citenamefont {Majer}, \citenamefont
  {Li}, \citenamefont {Kes},\ and\ \citenamefont
  {Konczykowski}}]{Khaykovich1996VL}%
  \BibitemOpen
  \bibfield  {author} {\bibinfo {author} {\bibfnamefont {B.}~\bibnamefont
  {Khaykovich}}, \bibinfo {author} {\bibfnamefont {E.}~\bibnamefont {Zeldov}},
  \bibinfo {author} {\bibfnamefont {D.}~\bibnamefont {Majer}}, \bibinfo
  {author} {\bibfnamefont {T.~W.}\ \bibnamefont {Li}}, \bibinfo {author}
  {\bibfnamefont {P.~H.}\ \bibnamefont {Kes}},\ and\ \bibinfo {author}
  {\bibfnamefont {M.}~\bibnamefont {Konczykowski}},\ }\bibfield  {title}
  {\bibinfo {title} {{Vortex-Lattice Phase Transitions in
  ${\mathrm{Bi}}_{2}{\mathrm{Sr}}_{2}{\mathrm{CaCu}}_{2}{\mathrm{O}}_{8}$
  Crystals with Different Oxygen Stoichiometry}},\ }\href
  {https://doi.org/10.1103/PhysRevLett.76.2555} {\bibfield  {journal} {\bibinfo
   {journal} {Phys. Rev. Lett.}\ }\textbf {\bibinfo {volume} {76}},\ \bibinfo
  {pages} {2555} (\bibinfo {year} {1996})}\BibitemShut {NoStop}%
\bibitem [{\citenamefont {Li}\ and\ \citenamefont {Wen}(2002)}]{li2002bg}%
  \BibitemOpen
  \bibfield  {author} {\bibinfo {author} {\bibfnamefont {S.}~\bibnamefont
  {Li}}\ and\ \bibinfo {author} {\bibfnamefont {H.-H.}\ \bibnamefont {Wen}},\
  }\bibfield  {title} {\bibinfo {title} {{No ending point on the Bragg glass
  phase transition line at low temperatures}},\ }\href
  {https://doi.org/10.1103/PhysRevB.65.214515} {\bibfield  {journal} {\bibinfo
  {journal} {Phys. Rev. B}\ }\textbf {\bibinfo {volume} {65}},\ \bibinfo
  {pages} {214515} (\bibinfo {year} {2002})}\BibitemShut {NoStop}%
\bibitem [{\citenamefont {Cubitt}\ \emph {et~al.}(1993)\citenamefont {Cubitt},
  \citenamefont {Forgan}, \citenamefont {Yang}, \citenamefont {Lee},
  \citenamefont {Paul}, \citenamefont {Mook}, \citenamefont {Yethiraj},
  \citenamefont {Kes}, \citenamefont {Li}, \citenamefont {Menovsky},
  \citenamefont {Tarnawski},\ and\ \citenamefont
  {Mortensen}}]{cubitt1993direct}%
  \BibitemOpen
  \bibfield  {author} {\bibinfo {author} {\bibfnamefont {R.}~\bibnamefont
  {Cubitt}}, \bibinfo {author} {\bibfnamefont {E.~M.}\ \bibnamefont {Forgan}},
  \bibinfo {author} {\bibfnamefont {G.}~\bibnamefont {Yang}}, \bibinfo {author}
  {\bibfnamefont {S.~L.}\ \bibnamefont {Lee}}, \bibinfo {author} {\bibfnamefont
  {D.~M.}\ \bibnamefont {Paul}}, \bibinfo {author} {\bibfnamefont {H.~A.}\
  \bibnamefont {Mook}}, \bibinfo {author} {\bibfnamefont {M.}~\bibnamefont
  {Yethiraj}}, \bibinfo {author} {\bibfnamefont {P.~H.}\ \bibnamefont {Kes}},
  \bibinfo {author} {\bibfnamefont {T.~W.}\ \bibnamefont {Li}}, \bibinfo
  {author} {\bibfnamefont {A.~A.}\ \bibnamefont {Menovsky}}, \bibinfo {author}
  {\bibfnamefont {Z.}~\bibnamefont {Tarnawski}},\ and\ \bibinfo {author}
  {\bibfnamefont {K.}~\bibnamefont {Mortensen}},\ }\bibfield  {title} {\bibinfo
  {title} {{Direct observation of magnetic flux lattice melting and
  decomposition in the high-$T_{\mathrm{c}}$ superconductor
  ${\mathrm{Bi}}_{2}{\mathrm{Sr}}_{2}{\mathrm{CaCu}}_{2}{\mathrm{O}}_{8}$}},\
  }\href {https://www.nature.com/articles/365407a0} {\bibfield  {journal}
  {\bibinfo  {journal} {Nature}\ }\textbf {\bibinfo {volume} {365}},\ \bibinfo
  {pages} {407} (\bibinfo {year} {1993})}\BibitemShut {NoStop}%
\bibitem [{\citenamefont {Nishizaki}\ \emph {et~al.}(1998)\citenamefont
  {Nishizaki}, \citenamefont {Naito},\ and\ \citenamefont
  {Kobayashi}}]{Nishizaki1998PhysRevB}%
  \BibitemOpen
  \bibfield  {author} {\bibinfo {author} {\bibfnamefont {T.}~\bibnamefont
  {Nishizaki}}, \bibinfo {author} {\bibfnamefont {T.}~\bibnamefont {Naito}},\
  and\ \bibinfo {author} {\bibfnamefont {N.}~\bibnamefont {Kobayashi}},\
  }\bibfield  {title} {\bibinfo {title} {{Anomalous magnetization and
  field-driven disordering transition of a vortex lattice in untwinned
  ${\mathrm{YBa}}_{2}{\mathrm{Cu}}_{3}{\mathrm{O}}_{y}$}},\ }\href
  {https://doi.org/10.1103/PhysRevB.58.11169} {\bibfield  {journal} {\bibinfo
  {journal} {Phys. Rev. B}\ }\textbf {\bibinfo {volume} {58}},\ \bibinfo
  {pages} {11169} (\bibinfo {year} {1998})}\BibitemShut {NoStop}%
\bibitem [{\citenamefont {Abulafia}\ \emph {et~al.}(1996)\citenamefont
  {Abulafia}, \citenamefont {Shaulov}, \citenamefont {Wolfus}, \citenamefont
  {Prozorov}, \citenamefont {Burlachkov}, \citenamefont {Yeshurun},
  \citenamefont {Majer}, \citenamefont {Zeldov}, \citenamefont {W\"uhl},
  \citenamefont {Geshkenbein},\ and\ \citenamefont
  {Vinokur}}]{Abulafia1996PhysRevLett}%
  \BibitemOpen
  \bibfield  {author} {\bibinfo {author} {\bibfnamefont {Y.}~\bibnamefont
  {Abulafia}}, \bibinfo {author} {\bibfnamefont {A.}~\bibnamefont {Shaulov}},
  \bibinfo {author} {\bibfnamefont {Y.}~\bibnamefont {Wolfus}}, \bibinfo
  {author} {\bibfnamefont {R.}~\bibnamefont {Prozorov}}, \bibinfo {author}
  {\bibfnamefont {L.}~\bibnamefont {Burlachkov}}, \bibinfo {author}
  {\bibfnamefont {Y.}~\bibnamefont {Yeshurun}}, \bibinfo {author}
  {\bibfnamefont {D.}~\bibnamefont {Majer}}, \bibinfo {author} {\bibfnamefont
  {E.}~\bibnamefont {Zeldov}}, \bibinfo {author} {\bibfnamefont
  {H.}~\bibnamefont {W\"uhl}}, \bibinfo {author} {\bibfnamefont {V.~B.}\
  \bibnamefont {Geshkenbein}},\ and\ \bibinfo {author} {\bibfnamefont {V.~M.}\
  \bibnamefont {Vinokur}},\ }\bibfield  {title} {\bibinfo {title} {{Plastic
  Vortex Creep in
  ${\mathrm{YBa}}_{2}{\mathrm{Cu}}_{3}{\mathrm{O}}_{7\ensuremath{-}x}$
  Crystals}},\ }\href {https://doi.org/10.1103/PhysRevLett.77.1596} {\bibfield
  {journal} {\bibinfo  {journal} {Phys. Rev. Lett.}\ }\textbf {\bibinfo
  {volume} {77}},\ \bibinfo {pages} {1596} (\bibinfo {year}
  {1996})}\BibitemShut {NoStop}%
\bibitem [{\citenamefont {Nishizaki}\ \emph {et~al.}(2000)\citenamefont
  {Nishizaki}, \citenamefont {Naito}, \citenamefont {Okayasu}, \citenamefont
  {Iwase},\ and\ \citenamefont {Kobayashi}}]{Nishizaki2000PhysRevB}%
  \BibitemOpen
  \bibfield  {author} {\bibinfo {author} {\bibfnamefont {T.}~\bibnamefont
  {Nishizaki}}, \bibinfo {author} {\bibfnamefont {T.}~\bibnamefont {Naito}},
  \bibinfo {author} {\bibfnamefont {S.}~\bibnamefont {Okayasu}}, \bibinfo
  {author} {\bibfnamefont {A.}~\bibnamefont {Iwase}},\ and\ \bibinfo {author}
  {\bibfnamefont {N.}~\bibnamefont {Kobayashi}},\ }\bibfield  {title} {\bibinfo
  {title} {{Effects of weak point disorder on the vortex matter phase diagram
  in untwinned ${\mathrm{YBa}}_{2}{\mathrm{Cu}}_{3}{\mathrm{O}}_{y}$ single
  crystals}},\ }\href {https://doi.org/10.1103/PhysRevB.61.3649} {\bibfield
  {journal} {\bibinfo  {journal} {Phys. Rev. B}\ }\textbf {\bibinfo {volume}
  {61}},\ \bibinfo {pages} {3649} (\bibinfo {year} {2000})}\BibitemShut
  {NoStop}%
\bibitem [{\citenamefont {Deligiannis}\ \emph {et~al.}(1997)\citenamefont
  {Deligiannis}, \citenamefont {de~Groot}, \citenamefont {Oussena},
  \citenamefont {Pinfold}, \citenamefont {Langan}, \citenamefont {Gagnon},\
  and\ \citenamefont {Taillefer}}]{Deligiannis1997PhysRevLett}%
  \BibitemOpen
  \bibfield  {author} {\bibinfo {author} {\bibfnamefont {K.}~\bibnamefont
  {Deligiannis}}, \bibinfo {author} {\bibfnamefont {P.~A.~J.}\ \bibnamefont
  {de~Groot}}, \bibinfo {author} {\bibfnamefont {M.}~\bibnamefont {Oussena}},
  \bibinfo {author} {\bibfnamefont {S.}~\bibnamefont {Pinfold}}, \bibinfo
  {author} {\bibfnamefont {R.}~\bibnamefont {Langan}}, \bibinfo {author}
  {\bibfnamefont {R.}~\bibnamefont {Gagnon}},\ and\ \bibinfo {author}
  {\bibfnamefont {L.}~\bibnamefont {Taillefer}},\ }\bibfield  {title} {\bibinfo
  {title} {{New Features in the Vortex Phase Diagram of
  ${\mathrm{YBa}}_{2}{\mathrm{Cu}}_{3}{\mathrm{O}}_{7\ensuremath{-}\mathit{\ensuremath{\delta}}}$}},\
  }\href {https://doi.org/10.1103/PhysRevLett.79.2121} {\bibfield  {journal}
  {\bibinfo  {journal} {Phys. Rev. Lett.}\ }\textbf {\bibinfo {volume} {79}},\
  \bibinfo {pages} {2121} (\bibinfo {year} {1997})}\BibitemShut {NoStop}%
\bibitem [{\citenamefont {Sarkar}\ \emph {et~al.}(2001)\citenamefont {Sarkar},
  \citenamefont {Pal}, \citenamefont {Paulose}, \citenamefont {Ramakrishnan},
  \citenamefont {Grover}, \citenamefont {Tomy}, \citenamefont {Dasgupta},
  \citenamefont {Sarma}, \citenamefont {Balakrishnan},\ and\ \citenamefont
  {Paul}}]{Sarkar2001PhysRevB}%
  \BibitemOpen
  \bibfield  {author} {\bibinfo {author} {\bibfnamefont {S.}~\bibnamefont
  {Sarkar}}, \bibinfo {author} {\bibfnamefont {D.}~\bibnamefont {Pal}},
  \bibinfo {author} {\bibfnamefont {P.~L.}\ \bibnamefont {Paulose}}, \bibinfo
  {author} {\bibfnamefont {S.}~\bibnamefont {Ramakrishnan}}, \bibinfo {author}
  {\bibfnamefont {A.~K.}\ \bibnamefont {Grover}}, \bibinfo {author}
  {\bibfnamefont {C.~V.}\ \bibnamefont {Tomy}}, \bibinfo {author}
  {\bibfnamefont {D.}~\bibnamefont {Dasgupta}}, \bibinfo {author}
  {\bibfnamefont {B.~K.}\ \bibnamefont {Sarma}}, \bibinfo {author}
  {\bibfnamefont {G.}~\bibnamefont {Balakrishnan}},\ and\ \bibinfo {author}
  {\bibfnamefont {D.~M.}\ \bibnamefont {Paul}},\ }\bibfield  {title} {\bibinfo
  {title} {{Multiple magnetization peaks in weakly pinned
  ${\mathrm{Ca}}_{3}{\mathrm{Rh}}_{4}{\mathrm{Sn}}_{13}$ and
  ${\mathrm{YBa}}_{2}{\mathrm{Cu}}_{3}{\mathrm{O}}_{7\ensuremath{-}\ensuremath{\delta}}$}},\
  }\href {https://doi.org/10.1103/PhysRevB.64.144510} {\bibfield  {journal}
  {\bibinfo  {journal} {Phys. Rev. B}\ }\textbf {\bibinfo {volume} {64}},\
  \bibinfo {pages} {144510} (\bibinfo {year} {2001})}\BibitemShut {NoStop}%
\bibitem [{\citenamefont {Pissas}\ \emph {et~al.}(2006)\citenamefont {Pissas},
  \citenamefont {Stamopoulos}, \citenamefont {Psycharis}, \citenamefont {Ma},\
  and\ \citenamefont {Wang}}]{Pissas2006PhysRevB}%
  \BibitemOpen
  \bibfield  {author} {\bibinfo {author} {\bibfnamefont {M.}~\bibnamefont
  {Pissas}}, \bibinfo {author} {\bibfnamefont {D.}~\bibnamefont {Stamopoulos}},
  \bibinfo {author} {\bibfnamefont {V.}~\bibnamefont {Psycharis}}, \bibinfo
  {author} {\bibfnamefont {Y.~C.}\ \bibnamefont {Ma}},\ and\ \bibinfo {author}
  {\bibfnamefont {N.~L.}\ \bibnamefont {Wang}},\ }\bibfield  {title} {\bibinfo
  {title} {{Splitting of the second magnetization peak in the superconductor
  ${\mathrm{Tl}}_{2}{\mathrm{Ba}}_{2}\mathrm{Ca}{\mathrm{Cu}}_{2}{\mathrm{O}}_{8+x}$}},\
  }\href {https://doi.org/10.1103/PhysRevB.73.174524} {\bibfield  {journal}
  {\bibinfo  {journal} {Phys. Rev. B}\ }\textbf {\bibinfo {volume} {73}},\
  \bibinfo {pages} {174524} (\bibinfo {year} {2006})}\BibitemShut {NoStop}%
\bibitem [{\citenamefont {Stamopoulos}\ and\ \citenamefont
  {Pissas}(2002)}]{Stamopoulos2002PhysRevB}%
  \BibitemOpen
  \bibfield  {author} {\bibinfo {author} {\bibfnamefont {D.}~\bibnamefont
  {Stamopoulos}}\ and\ \bibinfo {author} {\bibfnamefont {M.}~\bibnamefont
  {Pissas}},\ }\bibfield  {title} {\bibinfo {title} {{Hysteretic behavior of
  the vortex lattice at the onset of the second peak for the
  ${\mathrm{HgBa}}_{2}{\mathrm{CuO}}_{4+\ensuremath{\delta}}$
  superconductor}},\ }\href {https://doi.org/10.1103/PhysRevB.65.134524}
  {\bibfield  {journal} {\bibinfo  {journal} {Phys. Rev. B}\ }\textbf {\bibinfo
  {volume} {65}},\ \bibinfo {pages} {134524} (\bibinfo {year}
  {2002})}\BibitemShut {NoStop}%
\bibitem [{\citenamefont {Wang}\ \emph {et~al.}(2018)\citenamefont {Wang},
  \citenamefont {Luo}, \citenamefont {Li}, \citenamefont {Zeng}, \citenamefont
  {Cheng}, \citenamefont {Freyermuth}, \citenamefont {Tang}, \citenamefont
  {Yu}, \citenamefont {Yu}, \citenamefont {Greven},\ and\ \citenamefont
  {Li}}]{Wang2018PhysRevMaterials}%
  \BibitemOpen
  \bibfield  {author} {\bibinfo {author} {\bibfnamefont {L.}~\bibnamefont
  {Wang}}, \bibinfo {author} {\bibfnamefont {X.}~\bibnamefont {Luo}}, \bibinfo
  {author} {\bibfnamefont {J.}~\bibnamefont {Li}}, \bibinfo {author}
  {\bibfnamefont {J.}~\bibnamefont {Zeng}}, \bibinfo {author} {\bibfnamefont
  {M.}~\bibnamefont {Cheng}}, \bibinfo {author} {\bibfnamefont
  {J.}~\bibnamefont {Freyermuth}}, \bibinfo {author} {\bibfnamefont
  {Y.}~\bibnamefont {Tang}}, \bibinfo {author} {\bibfnamefont {B.}~\bibnamefont
  {Yu}}, \bibinfo {author} {\bibfnamefont {G.}~\bibnamefont {Yu}}, \bibinfo
  {author} {\bibfnamefont {M.}~\bibnamefont {Greven}},\ and\ \bibinfo {author}
  {\bibfnamefont {Y.}~\bibnamefont {Li}},\ }\bibfield  {title} {\bibinfo
  {title} {{Growth and characterization of
  ${\mathrm{HgBa}}_{2}{\mathrm{CaCu}}_{2}{\mathrm{O}}_{6+\ensuremath{\delta}}$
  and
  ${\mathrm{HgBa}}_{2}{\mathrm{Ca}}_{2}{\mathrm{Cu}}_{3}{\mathrm{O}}_{8+\ensuremath{\delta}}$
  crystals}},\ }\href {https://doi.org/10.1103/PhysRevMaterials.2.123401}
  {\bibfield  {journal} {\bibinfo  {journal} {Phys. Rev. Mater.}\ }\textbf
  {\bibinfo {volume} {2}},\ \bibinfo {pages} {123401} (\bibinfo {year}
  {2018})}\BibitemShut {NoStop}%
\bibitem [{\citenamefont {Bean}(1964)}]{BEAN1964RevModPhys}%
  \BibitemOpen
  \bibfield  {author} {\bibinfo {author} {\bibfnamefont {C.~P.}\ \bibnamefont
  {Bean}},\ }\bibfield  {title} {\bibinfo {title} {{Magnetization of High-Field
  Superconductors}},\ }\href {https://doi.org/10.1103/RevModPhys.36.31}
  {\bibfield  {journal} {\bibinfo  {journal} {Rev. Mod. Phys.}\ }\textbf
  {\bibinfo {volume} {36}},\ \bibinfo {pages} {31} (\bibinfo {year}
  {1964})}\BibitemShut {NoStop}%
\bibitem [{\citenamefont {Ovchinnikov}\ and\ \citenamefont
  {Ivlev}(1991)}]{OvchinnikovPhysRevB}%
  \BibitemOpen
  \bibfield  {author} {\bibinfo {author} {\bibfnamefont {Y.~N.}\ \bibnamefont
  {Ovchinnikov}}\ and\ \bibinfo {author} {\bibfnamefont {B.~I.}\ \bibnamefont
  {Ivlev}},\ }\bibfield  {title} {\bibinfo {title} {Pinning in layered
  inhomogeneous superconductors},\ }\href
  {https://doi.org/10.1103/PhysRevB.43.8024} {\bibfield  {journal} {\bibinfo
  {journal} {Phys. Rev. B}\ }\textbf {\bibinfo {volume} {43}},\ \bibinfo
  {pages} {8024} (\bibinfo {year} {1991})}\BibitemShut {NoStop}%
\bibitem [{\citenamefont {Xing}\ \emph {et~al.}(2020)\citenamefont {Xing},
  \citenamefont {Yi}, \citenamefont {Li}, \citenamefont {Meng}, \citenamefont
  {Mu}, \citenamefont {Ge},\ and\ \citenamefont {Shi}}]{xing2020vortex}%
  \BibitemOpen
  \bibfield  {author} {\bibinfo {author} {\bibfnamefont {X.}~\bibnamefont
  {Xing}}, \bibinfo {author} {\bibfnamefont {X.}~\bibnamefont {Yi}}, \bibinfo
  {author} {\bibfnamefont {M.}~\bibnamefont {Li}}, \bibinfo {author}
  {\bibfnamefont {Y.}~\bibnamefont {Meng}}, \bibinfo {author} {\bibfnamefont
  {G.}~\bibnamefont {Mu}}, \bibinfo {author} {\bibfnamefont {J.-Y.}\
  \bibnamefont {Ge}},\ and\ \bibinfo {author} {\bibfnamefont {Z.}~\bibnamefont
  {Shi}},\ }\bibfield  {title} {\bibinfo {title} {{Vortex phase diagram in
  12442-type
  ${\mathrm{RbCa}}_{2}{\mathrm{Fe}}_{4}{\mathrm{As}}_{4}{\mathrm{F}}_{2}$
  single crystal revealed by magneto-transport and magnetization
  measurements}},\ }\href
  {https://iopscience.iop.org/article/10.1088/1361-6668/abb35f/meta} {\bibfield
   {journal} {\bibinfo  {journal} {Supercond. Sci. Technol.}\ }\textbf
  {\bibinfo {volume} {33}},\ \bibinfo {pages} {114005} (\bibinfo {year}
  {2020})}\BibitemShut {NoStop}%
\bibitem [{\citenamefont {Cole}\ \emph {et~al.}(2023)\citenamefont {Cole},
  \citenamefont {Venuti}, \citenamefont {Gorman}, \citenamefont {Bauer},
  \citenamefont {Chan},\ and\ \citenamefont {Eley}}]{Cole2023PhysRevB}%
  \BibitemOpen
  \bibfield  {author} {\bibinfo {author} {\bibfnamefont {H.~M.}\ \bibnamefont
  {Cole}}, \bibinfo {author} {\bibfnamefont {M.~B.}\ \bibnamefont {Venuti}},
  \bibinfo {author} {\bibfnamefont {B.}~\bibnamefont {Gorman}}, \bibinfo
  {author} {\bibfnamefont {E.~D.}\ \bibnamefont {Bauer}}, \bibinfo {author}
  {\bibfnamefont {M.~K.}\ \bibnamefont {Chan}},\ and\ \bibinfo {author}
  {\bibfnamefont {S.}~\bibnamefont {Eley}},\ }\bibfield  {title} {\bibinfo
  {title} {{Plastic vortex creep and dimensional crossovers in the highly
  anisotropic superconductor ${\mathrm{HgBa}}_{2}{\mathrm{CuO}}_{4+x}$}},\
  }\href {https://doi.org/10.1103/PhysRevB.107.104509} {\bibfield  {journal}
  {\bibinfo  {journal} {Phys. Rev. B}\ }\textbf {\bibinfo {volume} {107}},\
  \bibinfo {pages} {104509} (\bibinfo {year} {2023})}\BibitemShut {NoStop}%
\bibitem [{\citenamefont {Giller}\ \emph {et~al.}(1999)\citenamefont {Giller},
  \citenamefont {Shaulov}, \citenamefont {Yeshurun},\ and\ \citenamefont
  {Giapintzakis}}]{Giller1999PhysRevB}%
  \BibitemOpen
  \bibfield  {author} {\bibinfo {author} {\bibfnamefont {D.}~\bibnamefont
  {Giller}}, \bibinfo {author} {\bibfnamefont {A.}~\bibnamefont {Shaulov}},
  \bibinfo {author} {\bibfnamefont {Y.}~\bibnamefont {Yeshurun}},\ and\
  \bibinfo {author} {\bibfnamefont {J.}~\bibnamefont {Giapintzakis}},\
  }\bibfield  {title} {\bibinfo {title} {{Vortex solid-solid phase transition
  in an untwinned
  ${\mathrm{YBa}}_{2}{\mathrm{Cu}}_{3}{\mathrm{O}}_{7\mathrm{\ensuremath{-}}\mathrm{\ensuremath{\delta}}}$
  crystal}},\ }\href {https://doi.org/10.1103/PhysRevB.60.106} {\bibfield
  {journal} {\bibinfo  {journal} {Phys. Rev. B}\ }\textbf {\bibinfo {volume}
  {60}},\ \bibinfo {pages} {106} (\bibinfo {year} {1999})}\BibitemShut
  {NoStop}%
\bibitem [{\citenamefont {Radzyner}\ \emph {et~al.}(2002)\citenamefont
  {Radzyner}, \citenamefont {Shaulov}, \citenamefont {Yeshurun}, \citenamefont
  {Felner}, \citenamefont {Kishio},\ and\ \citenamefont
  {Shimoyama}}]{Radzyner2002PhysRevB}%
  \BibitemOpen
  \bibfield  {author} {\bibinfo {author} {\bibfnamefont {Y.}~\bibnamefont
  {Radzyner}}, \bibinfo {author} {\bibfnamefont {A.}~\bibnamefont {Shaulov}},
  \bibinfo {author} {\bibfnamefont {Y.}~\bibnamefont {Yeshurun}}, \bibinfo
  {author} {\bibfnamefont {I.}~\bibnamefont {Felner}}, \bibinfo {author}
  {\bibfnamefont {K.}~\bibnamefont {Kishio}},\ and\ \bibinfo {author}
  {\bibfnamefont {J.}~\bibnamefont {Shimoyama}},\ }\bibfield  {title} {\bibinfo
  {title} {{Disorder and thermally driven vortex-lattice melting in
  ${\mathrm{La}}_{2\ensuremath{-}x}{\mathrm{Sr}}_{x}{\mathrm{CuO}}_{4}$
  crystals}},\ }\href {https://doi.org/10.1103/PhysRevB.65.100503} {\bibfield
  {journal} {\bibinfo  {journal} {Phys. Rev. B}\ }\textbf {\bibinfo {volume}
  {65}},\ \bibinfo {pages} {100503} (\bibinfo {year} {2002})}\BibitemShut
  {NoStop}%
\bibitem [{\citenamefont {Liu}\ \emph {et~al.}(2024)\citenamefont {Liu},
  \citenamefont {Xie},\ and\ \citenamefont {Wen}}]{Liu2024PhysRevB}%
  \BibitemOpen
  \bibfield  {author} {\bibinfo {author} {\bibfnamefont {Y.-H.}\ \bibnamefont
  {Liu}}, \bibinfo {author} {\bibfnamefont {W.}~\bibnamefont {Xie}},\ and\
  \bibinfo {author} {\bibfnamefont {H.-H.}\ \bibnamefont {Wen}},\ }\bibfield
  {title} {\bibinfo {title} {{Vanishing vortex creep at the transition from
  ordered to disordered vortex phases in
  ${\mathrm{Ba}}_{0.64}{\mathrm{K}}_{0.36}{\mathrm{Fe}}_{2}{\mathrm{As}}_{2}$}},\
  }\href {https://doi.org/10.1103/PhysRevB.109.214503} {\bibfield  {journal}
  {\bibinfo  {journal} {Phys. Rev. B}\ }\textbf {\bibinfo {volume} {109}},\
  \bibinfo {pages} {214503} (\bibinfo {year} {2024})}\BibitemShut {NoStop}%
\bibitem [{\citenamefont {Goffman}\ \emph {et~al.}(1998)\citenamefont
  {Goffman}, \citenamefont {Herbsommer}, \citenamefont {de~la Cruz},
  \citenamefont {Li},\ and\ \citenamefont {Kes}}]{Goffman1998PhysRevB}%
  \BibitemOpen
  \bibfield  {author} {\bibinfo {author} {\bibfnamefont {M.~F.}\ \bibnamefont
  {Goffman}}, \bibinfo {author} {\bibfnamefont {J.~A.}\ \bibnamefont
  {Herbsommer}}, \bibinfo {author} {\bibfnamefont {F.}~\bibnamefont {de~la
  Cruz}}, \bibinfo {author} {\bibfnamefont {T.~W.}\ \bibnamefont {Li}},\ and\
  \bibinfo {author} {\bibfnamefont {P.~H.}\ \bibnamefont {Kes}},\ }\bibfield
  {title} {\bibinfo {title} {{Vortex phase diagram of
  ${\mathrm{Bi}}_{2}{\mathrm{Sr}}_{2}{\mathrm{CaCu}}_{2}{\mathrm{O}}_{8+\ensuremath{\delta}}:$
  $c$-axis superconducting correlation in the different vortex phases}},\
  }\href {https://doi.org/10.1103/PhysRevB.57.3663} {\bibfield  {journal}
  {\bibinfo  {journal} {Phys. Rev. B}\ }\textbf {\bibinfo {volume} {57}},\
  \bibinfo {pages} {3663} (\bibinfo {year} {1998})}\BibitemShut {NoStop}%
\bibitem [{\citenamefont {Giamarchi}\ and\ \citenamefont
  {Le~Doussal}(1994)}]{Giamarchi1994PhysRevLett}%
  \BibitemOpen
  \bibfield  {author} {\bibinfo {author} {\bibfnamefont {T.}~\bibnamefont
  {Giamarchi}}\ and\ \bibinfo {author} {\bibfnamefont {P.}~\bibnamefont
  {Le~Doussal}},\ }\bibfield  {title} {\bibinfo {title} {Elastic theory of
  pinned flux lattices},\ }\href {https://doi.org/10.1103/PhysRevLett.72.1530}
  {\bibfield  {journal} {\bibinfo  {journal} {Phys. Rev. Lett.}\ }\textbf
  {\bibinfo {volume} {72}},\ \bibinfo {pages} {1530} (\bibinfo {year}
  {1994})}\BibitemShut {NoStop}%
\bibitem [{\citenamefont {Giamarchi}\ and\ \citenamefont
  {Le~Doussal}(1995)}]{Giamarchi1995PhysRevB}%
  \BibitemOpen
  \bibfield  {author} {\bibinfo {author} {\bibfnamefont {T.}~\bibnamefont
  {Giamarchi}}\ and\ \bibinfo {author} {\bibfnamefont {P.}~\bibnamefont
  {Le~Doussal}},\ }\bibfield  {title} {\bibinfo {title} {Elastic theory of flux
  lattices in the presence of weak disorder},\ }\href
  {https://doi.org/10.1103/PhysRevB.52.1242} {\bibfield  {journal} {\bibinfo
  {journal} {Phys. Rev. B}\ }\textbf {\bibinfo {volume} {52}},\ \bibinfo
  {pages} {1242} (\bibinfo {year} {1995})}\BibitemShut {NoStop}%
\bibitem [{\citenamefont {Koch}\ \emph {et~al.}(1989)\citenamefont {Koch},
  \citenamefont {Foglietti}, \citenamefont {Gallagher}, \citenamefont {Koren},
  \citenamefont {Gupta},\ and\ \citenamefont {Fisher}}]{Koch1989PhysRevLett}%
  \BibitemOpen
  \bibfield  {author} {\bibinfo {author} {\bibfnamefont {R.~H.}\ \bibnamefont
  {Koch}}, \bibinfo {author} {\bibfnamefont {V.}~\bibnamefont {Foglietti}},
  \bibinfo {author} {\bibfnamefont {W.~J.}\ \bibnamefont {Gallagher}}, \bibinfo
  {author} {\bibfnamefont {G.}~\bibnamefont {Koren}}, \bibinfo {author}
  {\bibfnamefont {A.}~\bibnamefont {Gupta}},\ and\ \bibinfo {author}
  {\bibfnamefont {M.~P.~A.}\ \bibnamefont {Fisher}},\ }\bibfield  {title}
  {\bibinfo {title} {{Experimental evidence for vortex-glass superconductivity
  in Y-Ba-Cu-O}},\ }\href {https://doi.org/10.1103/PhysRevLett.63.1511}
  {\bibfield  {journal} {\bibinfo  {journal} {Phys. Rev. Lett.}\ }\textbf
  {\bibinfo {volume} {63}},\ \bibinfo {pages} {1511} (\bibinfo {year}
  {1989})}\BibitemShut {NoStop}%
\bibitem [{\citenamefont {Kwok}\ \emph {et~al.}(2016)\citenamefont {Kwok},
  \citenamefont {Welp}, \citenamefont {Glatz}, \citenamefont {Koshelev},
  \citenamefont {Kihlstrom},\ and\ \citenamefont
  {Crabtree}}]{kwok2016vortices}%
  \BibitemOpen
  \bibfield  {author} {\bibinfo {author} {\bibfnamefont {W.-K.}\ \bibnamefont
  {Kwok}}, \bibinfo {author} {\bibfnamefont {U.}~\bibnamefont {Welp}}, \bibinfo
  {author} {\bibfnamefont {A.}~\bibnamefont {Glatz}}, \bibinfo {author}
  {\bibfnamefont {A.~E.}\ \bibnamefont {Koshelev}}, \bibinfo {author}
  {\bibfnamefont {K.~J.}\ \bibnamefont {Kihlstrom}},\ and\ \bibinfo {author}
  {\bibfnamefont {G.~W.}\ \bibnamefont {Crabtree}},\ }\bibfield  {title}
  {\bibinfo {title} {Vortices in high-performance high-temperature
  superconductors},\ }\href
  {https://iopscience.iop.org/article/10.1088/0034-4885/79/11/116501/meta}
  {\bibfield  {journal} {\bibinfo  {journal} {Rep. Prog. Phys}\ }\textbf
  {\bibinfo {volume} {79}},\ \bibinfo {pages} {116501} (\bibinfo {year}
  {2016})}\BibitemShut {NoStop}%
\end{thebibliography}%

\end{document}